\documentclass{pasj00}

\begin{document}
\SetRunningHead{Grechnev et al.}{2006 December 13 extreme solar
event}

\title{An Updated View of Solar Eruptive Flares and Development of Shocks and CMEs:
History of the 2006 December 13 GLE-Productive Extreme Event}

\author{Victor \textsc{Grechnev},
 Valentin \textsc{Kiselev},
 Arkadiy \textsc{Uralov},
 Nataliia \textsc{Meshalkina},
 Alexey \textsc{Kochanov}}
\affil{Institute of Solar-Terrestrial Physics SB RAS, Lermontov
St. 126A, Irkutsk 664033, Russia} \email{grechnev@iszf.irk.ru}



\maketitle

\begin{abstract}
An extreme 2006 December 13 event marked the onset of the Hinode
era being the last major flare in the solar cycle 23 observed with
NoRH and NoRP. The event produced a fast CME, strong shock, and
big particle event responsible for GLE70. We endeavor to clarify
relations between eruptions, shock wave, and the flare, and to
shed light on a debate over the origin of energetic protons. One
concept relates it with flare processes. Another one associates
acceleration of ions with a bow shock driven by a CME at
$(2-4)R_{\odot}$. The latter scenario is favored by a delayed
particle release time after the flare. However, our previous
studies have established that a shock wave is typically excited by
an impulsively erupting magnetic rope (future CME core) during the
flare rise, while the outer CME surface evolves from an arcade
whose expansion is driven from inside. Observations of the 2006
December 13 event reveal two shocks following each other, whose
excitation scenario contradicts the delayed CME-driven bow-shock
hypothesis. Actually, the shocks developed much earlier, and could
accelerate protons still before the flare peak. Then, the two
shocks merged into a single stronger one and only decelerated and
dampened long afterwards.
\end{abstract}

\hspace{2.5 truecm}
\parbox{7.5 truecm}
{\textit{Dedicated to the memory of T. Kosugi}}

\section{Introduction}
 \label{s-introduction}

Eruptions of solar magnetized plasmas and accompanying phenomena
present most vigorous manifestations of the solar activity.
General relations between eruptions and flares have been
theoretically understood in 1960--1970-s and constituted the basis
of the standard flare model (`CSHKP';
\cite{Car64,Sturrock66,Hirayama1974,Kopp76}). Associations of
eruptive flares with coronal shock waves were predicted by
\citet{Uchida1968} and \citet{Hirayama1974}. Later studies (e.g.,
\cite{Chen1989,Antiochos1999,Moore2001,Uralov2002}; and others)
supplemented the model with ideas about initiation of coronal mass
ejections (CMEs).

Despite the rather long history of theoretical concepts of solar
eruptive flares, difficulties still remain in establishing
connections between eruptions, flares, CMEs, and shock waves in
particular events. There is no consensus about such questions as
the excitation scenario of shock waves. Recent observations
promise noticeable update of model concepts. On the other hand,
progress in understanding the listed phenomena is currently urged
by requirements of modern industry, power, transport, and other
high-technology systems.

Eruptions and associated phenomena can produce severe space
weather disturbances. CMEs carry clouds of magnetized plasmas,
which can reach Earth and cause geomagnetic storms. Associated
shock waves also can affect space weather. Eruptions are
accompanied by flare emissions from radio waves up to gamma rays.
Solar eruptive events somehow accelerate electrons and protons to
high energies. Intense fluxes of accelerated protons sometimes
reach the Earth orbit being hazardous for equipment and astronauts
in space, while people on aircraft in high-latitude flights are
exposed to secondary particles.

The origin of solar energetic particles is still vague (see, e.g.,
\cite{Kallenrode2003}). There are two major competing concepts of
acceleration of high-energy heavy particles in solar events. One
concept relates their origin with flare processes within an active
region (e.g., \cite{KleinTrottet2001}). Another one relates
acceleration of protons and heavier ions up to high energies with
shock waves associated with CMEs (e.g.,
\cite{Cliver1982,Reames1999}). Specifying temporal properties of
the processes, which might be related to acceleration of protons,
is among important issues of both solar physics and space weather
forecast.

In the widely accepted conjectural scenario of particle
acceleration by a shock front, the shock wave is supposed to
develop as a bow shock driven by the outer surface of a
super-Alfv{\'e}nic CME (e.g., \cite{Reames2009,Aschwanden2012}).
One of arguments in favor of this scenario is an apparent delay of
the extrapolated solar particle release time relative to flare
emissions. This circumstance is considered `\textit{as further
evidence that the particles are accelerated by the shock wave that
forms late in the event, when the CME driver of the shock reaches
2 or 3 solar radii}' [heliocentric distance, or a height above the
solar surface of $(1-2)R_{\odot}$, \cite{Reames2009}]. The concept
of the bow shock originates from the analogy with the problem of
the supersonic plasma flow around the surface of a solid or
elastic body associated with the magnetic bubble of a CME. Two
circumstances are important here: (i)~the flow around the body
occurs with the existence of a stagnation point of the plasma flow
at the surface of the body, and (ii)~the motion velocity of this
stagnation point exceeds the ambient fast-mode speed. However, at
the CME formation stage, the analogy with such a plasma flow does
not apply, even though this flow is a subsonic one. At this stage,
the CME magnetic bubble extrudes surrounding magnetoplasmas away
almost omnidirectionally, thus forming an extensive disturbed zone
of compression around it. This zone is comparable in size with the
initial CME. The front of this zone is a weak discontinuity
running with the ambient fast-mode speed. The fast-mode speed of
the moving plasmas within this zone is higher than the pre-event
ambient fast-mode speed. The boundary of the CME magnetic bubble,
i.e., its outer separatrix surface, is already detectable at this
stage due to the growing plasma compression ahead. In the
bow-shock concept, the CME size and speed ($V_{\mathrm{CME}} >
V_{\mathrm{fast}}$) determine the position and intensity of the
stationary shock ahead of the CME. Kinematical differences of the
structural CME components preceding the appearance of the shock
are not discussed in the bow-shock concept. A significant
distinction of their accelerations from the self-similar regime is
among these differences.

However, the results of \authorcite{Grechnev2011_I}
(\yearcite{Grechnev2011_I,Grechnev2011_III}) do not support the
bow-shock excitation scenario in the low corona. Instead, it turns
out that shock waves are excited by the impulsive-piston mechanism
during the rise phase of a flare. In the impulsive-piston concept,
the wave disturbance is essentially non-stationary. Its intensity
is determined by the acceleration of the piston. The major role of
the acceleration can be demonstrated in the following way. The
magnetic flux rope expands in both the major and minor radii
simultaneously. Accordingly, the radiation of the magnetosonic
wave by an element of the flux rope can be divided into the dipole
and monopole components. The intensity of each component is
proportional to the squared acceleration, with which each radius
changes. The sharpest portion in the velocity profile of the
disturbance propagating away from the piston forms approximately
at the same time as the acceleration reaches its maximum. This
portion is a place, where the discontinuity (i.e., shock) starts
to form. As our previous analyses have shown, the wave front,
which appears in the disturbed zone surrounding a CME, is excited
by a sharp impulsive eruption inside the developing CME, where a
steep outward-directed falloff of the Alfv{\'e}n speed favors
amplification of the wave and rapid formation of the discontinuity
in $\sim 10^2$~s \citep{AfanasyevUralovGrechnev2013}. The
kinematics of the whole CME determines whether or not the heading
portion of the wave transforms into the bow shock afterwards.

This scenario is confirmed by \cite{Grechnev2011_I}, who briefly
discussed moderate eruptive flares. It is reasonable to check what
occurred in a major event, which has produced a big near-Earth
enhancement of high-energy proton flux. The extreme 2006 December
13 solar event observed in detail with many instruments provides
this opportunity.

Many papers already addressed various aspects of this event. At
least, two eruptive episodes have been revealed (e.g.,
\cite{Asai2008,Sterling2011}). The event produced large-scale
disturbances such as `EUV waves' and dimmings, the latter being
both deep depressions near the active region and shallower remote
dimmings (e.g., \cite{Attrill2010}). \citet{Liu2008} followed the
related CME and shock wave to the Earth orbit and then up to
2.7~AU. Several studies addressed the flare (e.g.,
\cite{Jing2008,Ning2008}). Some attempts have already been made to
find out the origin of near-Earth protons (e.g., \cite{Li2009};
\cite{Reames2009};
\authorcite{Firoz2011} \yearcite{Firoz2011,Firoz2012}).

Nevertheless, some important questions remain unanswered. We are
not aware of kinematic measurements of eruptions. Relations
between the eruptions and extreme flare have not been revealed.
The origin and onset time of the shock wave still remains
hypothetical. Some conclusions do not stand against observations
or contradict each other. For example, it is difficult to
reconcile the conclusion of \citet{Sterling2011} that the major
eruption occurred away from the strong fields with the result of
\citet{Jing2008} that the high energy release regions tend to be
concentrated in local strong field regions. There is a
contradiction between the conclusions of \citet{Li2009}, who
argued in favor of flare-acceleration of solar energetic
particles, and the conclusions of
\authorcite{Firoz2011} (\yearcite{Firoz2011,Firoz2012}), who favor
shock-acceleration of GLE particles.

In the present study we endeavor to shed further light on the
listed issues based on the approaches and techniques developed by
\citet{Grechnev2011_I} and briefly described in
section~\ref{s-techniques}. To reach the purposes listed above, we
firstly reveal in section~\ref{s-time_profiles} the features of
the major phase of the event from its time profiles. In
section~\ref{s-eruptions}, we measure the kinematical
characteristics of eruptions from the images observed with the
X-Ray Telescope (XRT; \cite{Golub2007}) and the Solar Optical
Telescope (SOT; \cite{Suematsu2008,Tsuneta2008}) on Hinode
\citep{Kosugi2007}. We then co-ordinate the eruptions with the
milestones of the extreme flare shown by the microwave total flux
light curves recorded with the Nobeyama Radio Polarimeters (NoRP;
\cite{Torii1979,Nakajima1985}). Section~\ref{s-flare} considers
the development of the extreme flare from microwave images
produced with the Nobeyama Radioheliograph (NoRH;
\cite{Nakajima1994}) along with NoRP, XRT, and SOT observations
starting from the early onset of the flare and up to the end of
the second flare peak. In this way, we address the circumstances
responsible for the extreme properties of the event.
Section~\ref{s-euv_wave} reveals near-surface traces of two shock
waves following each other, whose onset times and positions
quantitatively indicate their excitation by two major eruptions.
Section~\ref{s-dynamic_spectrum} confirms this result by the
analysis of the type II bursts in the dynamic radio spectrum.
Section~\ref{s-cme} discusses the CME, manifestations of the shock
in its structure, and their correspondence with the shock waves
revealed in previous sections. Section~\ref{s-discussion}
coordinates the results into a consistent picture of the whole
event and addresses some of the contradicting conclusions drawn
previously. The outcome of the analysis and its implications are
summarized in section~\ref{s-conclusion}.

\section{Measurement Techniques}
 \label{s-techniques}

Imaging observations provide important visual information about
eruptions, wavelike disturbances, and flares. In addition,
quantitative kinematical measurements of moving features can shed
light on causal relations between the listed phenomena. Such
measurements are complicated by difficulties to follow an
expanding feature in question due to its rapidly decreasing
brightness or opacity, while concurrent flare emissions are very
bright. Difficulties to detect an eruptive feature in all images
of interest lead to large uncertainties in its position. A
traditional way to measure velocities and accelerations by
differentiation of distance-time measurements causes large scatter
of results. To overcome this difficulty, we describe the
kinematics of eruptions and wavelike disturbances by analytic
functions and calculate the kinematical plots by means of
integration or differentiation of the analytic fit rather than the
measurements. The distance-time measurements are used as a
starting estimate of kinematical parameters, and then these
parameters are adjusted to outline the measured feature in a best
way. Our ultimate criterion is to follow the motion of an analyzed
feature in images as closely as possible.

\subsection{Motions of Eruptions}

Several studies (e.g., \cite{Zhang2001}; \cite{Maricic2007};
\authorcite{Temmer2008} \yearcite{Temmer2008,Temmer2010};
\authorcite{Grechnev2011_I} \yearcite{Grechnev2011_I,Grechnev2013})
have concluded that the acceleration of an eruption or a CME
occurs impulsively and temporally close to an associated HXR
burst. Using this fact, we consider the initial $v_0$ and final
$v_1$ velocities of an eruption to be nearly constant and fit its
acceleration with a Gaussian, $ a = \left( v_1-v_0 \right)
\exp{\{-{[(t-t_0)/\tau_{\mathrm{acc}}]^2}/2 \}} /
(\sqrt{2\pi}\tau_{\mathrm{acc}})$. Here
$\tau_{\mathrm{acc}}\sqrt{8\ln{2}}$ is the FWHM of the
acceleration time profile centered at time $t_0$. In cases of more
complex kinematics, we use a combination of Gaussians and adjust
their parameters manually.

With rather accurately estimated effective duration and center
time of the acceleration, its actual shape is rather uncertain
because of double integration in calculating the height-time plot,
which is directly compared with the observations. However, the
acceleration plot is not expected to contain features shorter than
the Alfv{\' e}n time inside the measured eruption.

\subsection{Waves}
 \label{S-wave_expansion}

A simple model (\authorcite{Grechnev2008waves}
\yearcite{Grechnev2008waves,Grechnev2011_I,Grechnev2011_III})
describes propagation of impulsively excited shock waves in plasma
with a radial power-law density falloff $\delta$ from an eruption
center, $n = n_0(x/h_0)^{-\delta}$. Here $x$ is the distance, and
$n_0$ is the density at a distance of $h_0 \approx 100$ Mm (close
to the scale height). Propagation of the global front of such a
shock wave is almost insensitive to the magnetic field, but is
determined by the plasma density distribution, $x(t) \propto
t^{2/(5-\delta)}$. This equation is more convenient to use in a
form $x(t) = x_1[(t-t_0)/(t-t_1)]^{2/(5-\delta)}$, where $t$ and
$x$ are the current time and distance, $t_0$ is the wave onset
time, and $(t_1,\ x_1)$ correspond to one of the measured fronts.

For the shock propagation along the solar surface, this simple
approximation also provides close results to those produced with
the analytic modeling of weak shock waves
\citep{AfanasyevUralov2011,Grechnev2011_III}. We use the same
approximation to fit the expansion of shock-associated CME
components as well as the drift rate of type II bursts.

\section{Observations}

\subsection{Time Profiles}
 \label{s-time_profiles}

\begin{figure}
  \begin{center}
    \FigureFile(85mm,130mm){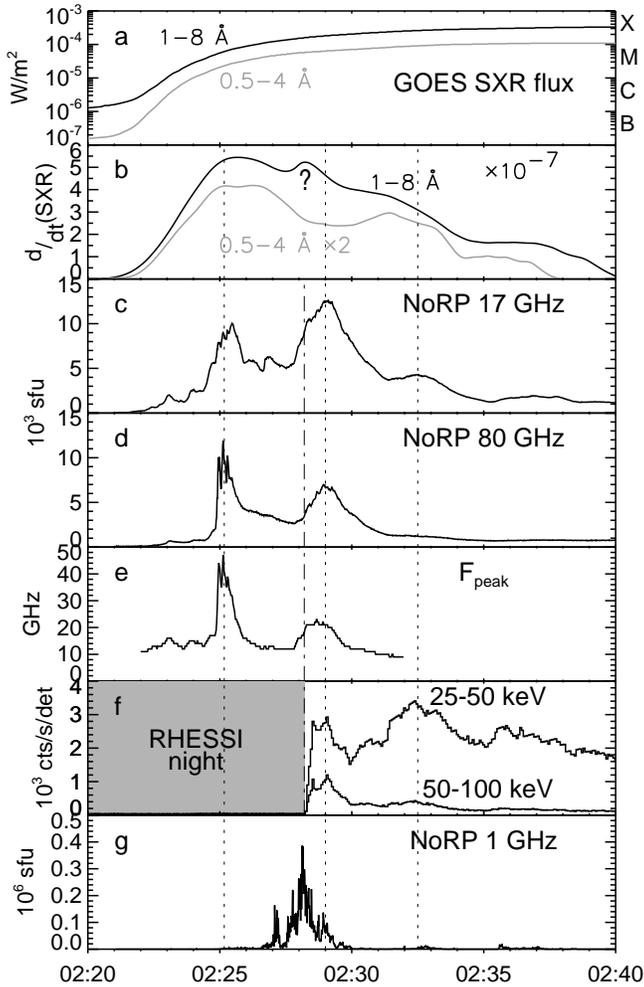}
  \end{center}
  \caption{Time profiles of the event. a)~GOES SXR flux records at 1--8~\AA\
(black) and 0.5--4~\AA\ (gray) and their derivatives (b; the peak
indicated by the question mark is probably an artifact); NoRP
total flux records at 17~GHz (c), 80~GHz (d), and the microwave
peak frequency (e); RHESSI HXR records at 25--50 keV and 50--100
keV (f), and a huge decimetric burst at 1 GHz (g, NoRP). The
shaded interval in panel (f) and its dash-dotted continuation
indicate RHESSI nighttime. Three vertical dotted lines mark the
highest peaks at 17 GHz.}
    \label{fig:timeprof}
 \end{figure}

Figure~\ref{fig:timeprof} characterizes the overall course of the
event. The soft X-ray (SXR) flux of the flare reached the X3.4
level (Figure~\ref{fig:timeprof}a). Two major flare peaks at 02:25
and 02:29 are conspicuous at 17 and 80 GHz
(Figure~\ref{fig:timeprof}c,d). Weaker quasi-periodic peaks with
an interval of $\sim 5$~min pronounced at lower frequencies
continued afterwards for more than one hour.

Figure~\ref{fig:timeprof}b shows the derivatives of the two SXR
GOES channels (splined 3-sec data). The derivative of the
1--8~\AA\ channel is inconclusive, because a subsidiary peak (the
question mark) is most likely an artifact due to the discrete
`staircase' time profile, which started at that time. The second
peak at 02:29 is obviously absent in the 0.5--4~\AA\ channel thus
deviating from the Neupert effect \citep{Neupert1968}. This issue
will be addressed in section~\ref{s-neupert_effect}.

Figure~\ref{fig:timeprof}e presents evolution of the microwave
peak (turnover) frequency computed  from NoRP total flux data by
using the second-order fit of instant log-log spectra averaged
over 1.2~s (see \cite{White2003,Grechnev2008protons}). The
turnover frequency reaches very high values, exceeding 35 GHz
during the first peak and 17 GHz during the second peak. According
to expressions of \citet{DulkMarsh1982}, this fact along with very
high flux densities $> 10^4$~sfu observed at 17 and 80 GHz
indicates emission from very large number of high-energy electrons
in strongest magnetic fields. Flaring during the first peak most
likely was stronger and harder than during the second one. The
microwave-emitting source was certainly optically thick at both 17
and 34 GHz during the first peak and at 17 GHz during the second
peak. However, the turnover frequency of $\leq 20$~GHz during the
second peak does not guarantee that the 34 GHz source was
optically thin at that time \citep{Kundu2009}.

RHESSI missed the first peak and the onset of the second peak due
to nighttime (Figure~\ref{fig:timeprof}f). Subsequent evolution of
the hard X-ray (HXR) emission in the 25--50 and 50--100 keV ranges
shows progressive softening of the HXR spectrum supporting the
assumption that the missed first peak could be still harder.

A huge decimetric burst reached almost half a million sfu at 1~GHz
(Figure~\ref{fig:timeprof}g). Its enormous intensity and a sharp
spiky time profile indicate an underlying coherent emission
mechanism interpreted by \citet{Kintner2009} as electron-cyclotron
maser (ECM). This burst was only superseded by a burst on December
6 from the same active region 10930. These huge radio bursts
caused failures of the GPS and GLONASS navigation systems
(\authorcite{Afraimovich2009a}
\yearcite{Afraimovich2009a,Afraimovich2009b}; \cite{Kintner2009}).

\subsection{Eruptions}
 \label{s-eruptions}

Eruptions in the 2006 December 13 event were previously considered
by \citet{Asai2008} (eruptions EF2 and EF3 in our notation),
\citet{Sterling2011} (eruption EF2), and \citet{Kusano2012}
(eruption EF1). Nevertheless, their kinematics has not been studied
so far. We analyzed the eruptions from SXR Hinode/XRT images. Their
motions were measured in the same way as
\authorcite{Grechnev2011_I}
(\yearcite{Grechnev2011_I,Grechnev2013}) did (see
section~\ref{s-techniques}). We firstly measured the positions of
the leading edge of each eruptive feature from the images. The
measured points were used as starting estimates to find the
initial and final velocities. Then we endeavored to reproduce the
motion of a feature in question by describing its acceleration
time profile of a Gaussian shape. The kinematic parameters were
adjusted in sequential attempts to follow the motion of a feature
in question as closely as possible. The motion of the second
eruption was more complex: its acceleration was immediately
followed by strong deceleration. We used its acceleration profile
as a combination of two Gaussian curves and adjusted their
parameters manually.

\begin{figure}
  \begin{center}
    \FigureFile(85mm,81mm){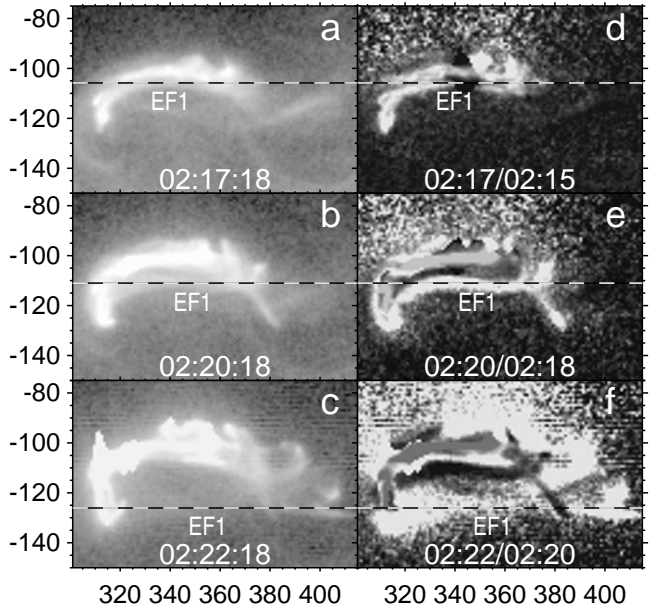}
  \end{center}
  \caption{Hinode/XRT images of eruptive feature EF1: direct images
(left column) and image ratios (right column, the times for each
pair of the images are indicated in panels d--e). The position of
the horizontal dashed line across the southernmost edge of EF1
corresponds to the kinematical measurements presented in
Figure~\ref{fig:accelerations}a with the solid curve. The axes
show hereafter in similar images arc seconds from the solar disk
center.}
 \label{fig:eruption1}
 \end{figure}

\begin{figure}
  \begin{center}
    \FigureFile(85mm,57mm){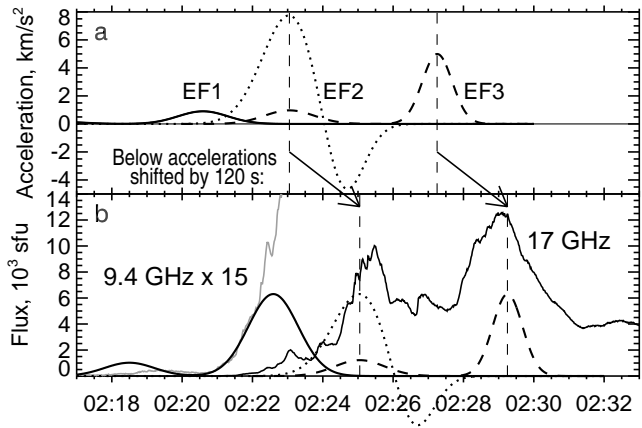}
  \end{center}
  \caption{(a)~Measured acceleration plots of eruptive features EF1
(solid), EF2 (dotted), and EF3 (dashed) and (b)~microwave time
profiles at 9.4 GHz (gray, magnified by a factor of 15) and at 17
GHz (solid) together with delayed normalized acceleration profiles
of EF1--EF3. The acceleration plots in panel (b) are delayed by
120~s. The vertical dashed lines denotes the acceleration peaks of
EF2 and EF3.}\label{fig:accelerations}
 \end{figure}

Hinode/XRT images reveal three eruptive features following each
other. Figure~\ref{fig:eruption1} presents the first eruptive
feature EF1 (left: direct images, right: image ratios), which was
weakly visible as a faint elongated brightening extended
East--West. EF1 separated from a bright bundle of loops and moved
south in the plane of the sky. As \citet{Kusano2012} showed, the
initial position of this eruptive feature coincided with the
magnetic polarity inversion (neutral) line, which is highlighted
in Figures~\ref{fig:eruption1}a--c by the brightest loop-like
bundle arranged along it. The displacement from the initial
position of the horizontal dashed line crossing the southernmost
bend of EF1 was calculated from the solid acceleration plot in
Figure~\ref{fig:accelerations}a. This feature probably was a
magnetic flux rope structure, as its initial position along the
neutral line implies. The acceleration of EF1 reached in the plane
of the sky 1~km~s$^{-2}$ at 02:20:30, and its speed reached
110~km~s$^{-1}$, while EF1 was visible.

\begin{figure*}
  \begin{center}
    \FigureFile(170mm,34mm){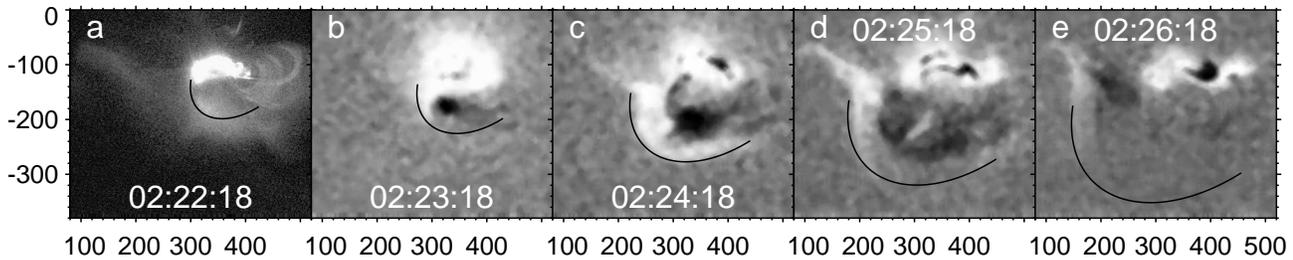}
  \end{center}
  \caption{Hinode/XRT images of the eruptive feature EF2.
Panel (a) presents an original image without subtraction. Images
in panels (b--e) are running differences. The black ovals outlines
the leading upper part of EF2 according to the dotted acceleration
plot in Figure~\ref{fig:accelerations}a.}
 \label{fig:eruption2}
 \end{figure*}

The second eruptive feature EF2 is shown in
Figure~\ref{fig:eruption2}. In the course of expansion, this
feature resembled a bow for shooting arrows; its east part only is
clearly visible. The acceleration plot of EF2 is presented with
the dotted curve in Figure~\ref{fig:accelerations}a. EF2
accelerated up to 8~km~s$^{-2}$ at 02:23, reached 750~km~s$^{-1}$,
and then strongly decelerated to 456~km~s$^{-1}$. \citet{Asai2008}
provided a close estimate of its plane-of-sky speed of
650~km~s$^{-1}$. They also revealed a strong blue shift of
emission whose hot ($> 2$~MK) source was close to this feature
(`BS2' in their notation) from spectroscopic data of Hinode/EIS at
02:22--02:24. The authors considered EF2 as a manifestation of an
MHD shock wave.

However, properties of feature EF2 are inconsistent with its wave
nature. \citet{Sterling2011} considered feature EF2 as a magnetic
loop system pushed outward by some core eruption, so that the
initial position of EF2 might have coincided with a static loop in
Figure~\ref{fig:eruption2}a. Nevertheless, excitation of shock
waves in this event conjectured by \citet{Asai2008} is undoubted
as shown by, e.g., \citet{Liu2008}, and we will confirm this
later.

\begin{figure}
  \begin{center}
    \FigureFile(80mm,128mm){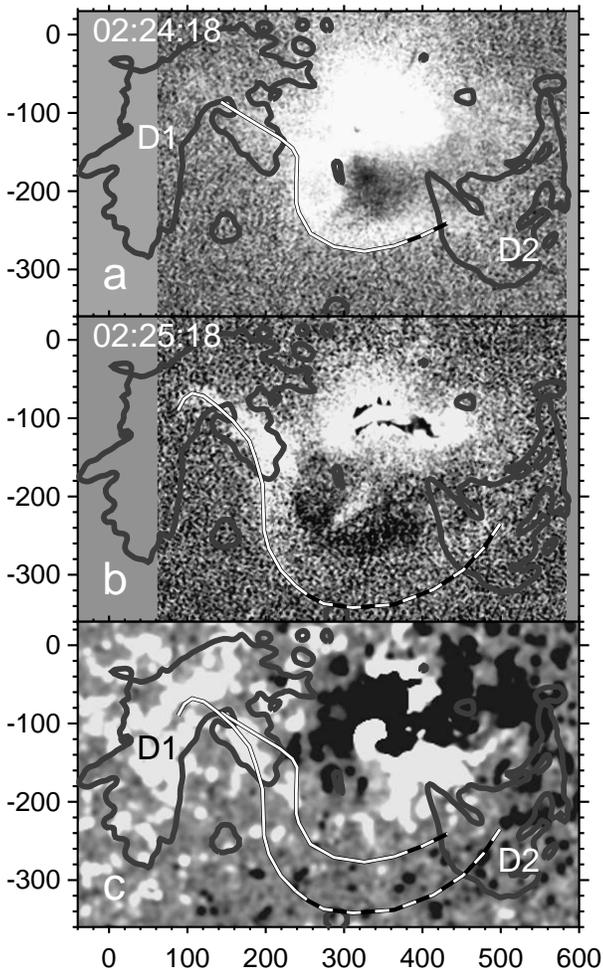}
  \end{center}
  \caption{Two XRT images of the eruptive feature EF2 (a,b)
in comparison with double dimming regions D1 and D2 (dark gray
contours) and the MDI magnetogram (c). The solid curves outline the
parts of EF2 clearly visible in these images. The dashed parts of
the curves outline suggestions of EF2 revealed from the XRT images
processed in different ways.}\label{fig:eruption2_dimming}
 \end{figure}

Before comparison of the kinematical properties of the eruptive
features EF1, EF2, and EF3 with different observational facts, now
we try to understand what feature EF2 could be in nature. This
eruptive feature was the largest in size and most impulsive in
this event. The eruption has resulted in the appearance of the
double major regions of coronal dimming on the periphery of
AR~10930 (see \authorcite{Imada2007}
\yearcite{Imada2007,Imada2011}; \cite{Jin2009,Attrill2010}).
Figure~\ref{fig:eruption2_dimming} presents the eruption of EF2,
the dimming regions, and the SOHO/MDI magnetogram produced at
01:40, short before the event. The dimming regions were revealed
from a difference of the SOHO/EIT 195~\AA\ images observed after
the event and before it. The criterion to select dimming was a
brightness decrease by 25 counts/pixel (the quiet Sun's level was
40 counts/pixel). To eliminate complicating small-scale structural
features, both the EIT difference image and the magnetogram were
smoothed by convolution with a two-dimensional Gaussian kernel (4
pixels width).

The solid curves in Figure~\ref{fig:eruption2_dimming} outline the
clearly visible parts of EF2; the dashed curves outline their
possible extensions (cf. Figure~\ref{fig:eruption2}). Comparison
of Figures \ref{fig:eruption2_dimming} and \ref{fig:eruption2}
shows that the eastern end of the `bow' was fixed and located
within the large dimming region D1, whose magnetic polarity was
positive (Figure~\ref{fig:eruption2_dimming}c). Thus, the western
end of the bow-like feature EF2 must be rooted in a
negative-polarity region. As the outlining curves indicate, the
western end of EF2 was most likely located in the dimming region
D2 dominated by the negative polarity.

In classical `double dimming' events, coronal dimming regions are
considered as the opposite-polarity footpoints of the ejected
CME's flux rope
\citep{HudsonWebb1997,SterlingHudson1997,Webb2000,Mandrini2005}.
This concept appears to be consistent with the discussed
properties of feature EF2, which was probably an eruptive flux
rope, the largest one and most impulsive in this event. The
magnetic fields enveloping the progenitor of the flux rope from
above in the strongest-field part of AR~10930 were directed
northward, and the axial field was directed westward, as
Figure~\ref{fig:eruption2_dimming}c shows. This arrangement of the
magnetic polarities agrees with the left handedness of the active
region indicated by its mirrored-S-shaped configuration and the
negative helicity of the corresponding near-Earth magnetic cloud
\citep{Liu2008}.

The progression of reconnection caused by the eruption formed the
flare arcade, whose development involved strongest magnetic fields
between the sunspots. The process seems to be well described by
the standard flare model. A seemingly contradiction with the
conclusion of \citet{Sterling2011} about the major eruption aside
of strong fields (their Figures 6, 8 and 9) will be reconciled in
section~\ref{s-cme}.

\begin{figure*}
  \begin{center}
    \FigureFile(170mm,35mm){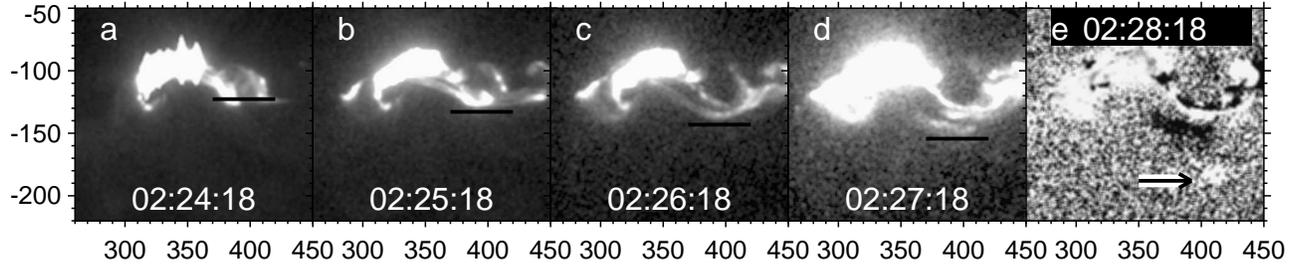}
  \end{center}
  \caption{Hinode/XRT images of the eruptive feature EF3. Panels (a--d)
present non-subtracted images, and panel (e) shows a difference
image. The black bars in panels (a--d) mark the farthest bend of
EF3 according to the dashed acceleration plot in
Figure~\ref{fig:accelerations}a. The black arrow in panel (e)
points at a faint latest manifestation of EF3.}
 \label{fig:eruption3}
 \end{figure*}

The second eruptive episode probably inspired eruption of the
third feature EF3 presented in Figure~\ref{fig:eruption3}. Its
acceleration plot is shown with a dashed curve in
Figure~\ref{fig:accelerations}a. Feature EF3 was initially located
along the neutral line; probably, it was also a flux rope. EF3
underwent two acceleration episodes. The first, a weaker one,
occurred simultaneously with acceleration of EF2 at 02:23. This
determined its appearance simultaneously with the first microwave
peak. The maximum acceleration of EF3 reached 5~km~s$^{-2}$ at
02:27 with a final speed of 420~km~s$^{-1}$. \citet{Asai2008}
revealed a strong blue shift for this feature also (BS1 in their
notation) and interpreted it as an ejected plasmoid. The
plane-of-sky speed estimated by the authors, $90$~km~s$^{-1}$, was
less than our measurements show (probably due to difficulties to
reveal its faint latest manifestation at 02:28:18 in
Figure~\ref{fig:eruption3}e), while its line-of-sight speed of
$240-280$~km~s$^{-1}$ estimated by \citet{Asai2008} seems to agree
with our result.

From the fact that the appearance of EF3 corresponded to the first
microwave peak recorded with NoRP \citet{Asai2008} reasonably
concluded that the flare was a product of magnetic reconnection
induced by the eruption. Comparison of the detailed acceleration
plots for eruptive features EF1, EF2, and EF3 in
Figure~\ref{fig:accelerations}a with microwave bursts in
Figure~\ref{fig:accelerations}b reveals a relation between them,
which appears to be still more impressive.

Figure~\ref{fig:accelerations}b presents the microwave time
profiles at 17 GHz (black solid) and at 9.4 GHz (gray, a magnified
initial part) along with normalized acceleration plots from
Figure~\ref{fig:accelerations}a shown with the same line styles
and arbitrarily shifted by 2~min later. The early rise of the 9.4
GHz emission coincides with the delayed acceleration of EF1, and
the 17 GHz time profile exhibits a striking similarity with the
delayed acceleration curves of EF2 and EF3. Thus, the flare bursts
were caused by the eruptions and not vice versa. Each eruption
apparently was causally related with another and preceded a flare
burst.

This relation is consistent with the standard flare model,
especially in its modern form. \citet{Kusano2012} numerically
simulated MHD processes caused by a wide variety of magnetic
structures and compared the results with Hinode imaging data.
According to their conclusion, the trigger of the 2006 December 13
flare manifested in the eruption of EF1; this is consistent with
the measured temporal relation between the acceleration of EF1 and
the microwave flare emission presented in our
Figure~\ref{fig:accelerations}.

\begin{figure}
  \begin{center}
    \FigureFile(85mm,87mm){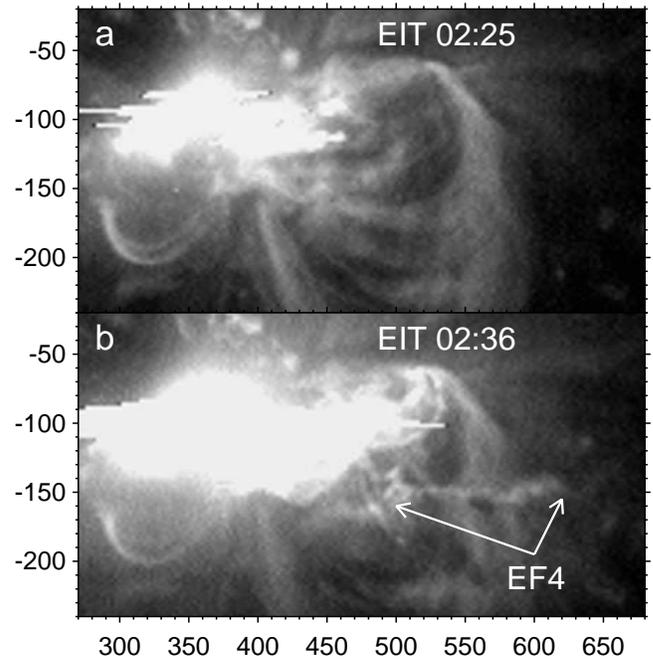}
  \end{center}
  \caption{Eruptive feature EF4 in EIT 195~\AA\ image at 02:36
indicated by the arrows (b). EF4 is certainly absent in the
preceding EIT 195~\AA\ image at 02:25.}
  \label{fig:eruption4}
 \end{figure}

Eruption EF3 was not the last one in this event.
Figure~\ref{fig:eruption4}b presents one more eruptive feature EF4
in a SOHO/EIT 195~\AA\ image mentioned by \citet{Asai2008}. The
position of this feature at 02:36 rules out its identity with any
of the preceding ones. This feature is not visible in either
Hinode/XRT images (probably its temperature was lower than XRT can
see) or GOES-12/SXI ones. It is not possible to measure its motion
from a single image. Assuming that its relation to microwaves was
approximately the same as for the preceding eruptions, one might
assume the onset of its motion at the early rise of the next
microwave peak marked by the dotted line in
Figure~\ref{fig:timeprof} (02:32:30) or about 2 min before, i.e.,
at 02:29--02:31. The initial position of EF4 might be marked by
the westernmost loops in XRT images and ribbons in SOT ones in
Figure~\ref{fig:norh_hinode}. With this assumption, the
plane-of-sky speed of its leading edge should be
250--400~km~s$^{-1}$, which seems to be reasonable. Thus,
association of the eruptive feature EF4 with the next, smaller
flare peak at 02:32:30 is possible.

The time profiles of accelerations in
Figure~\ref{fig:accelerations} had a quasi-periodic character.
Eruption of EF4 probably followed this trend. We remind that the
quasi-periodic pulsations continued long afterwards in microwaves.
This circumstance induces thinking about possible causes of the
oscillatory behavior of this event and the role of the
oscillations in triggering the eruptions. However, this issue is
beyond our scope.

\subsection{Development of the Flare}
 \label{s-flare}
 \subsubsection{Pre-flare Emission}
\citet{Smolkov2009} have shown that microwave emission of active
region 10930 at 17 GHz in relatively quiet conditions during a
week before the December 13 flare was dominated by a neutral line
source (NLS). Such long-lived sources reside in the vicinity of
the main neutral line where the horizontal magnetic component is
maximum \citep{Uralov2006}. Emission of a NLS at 17 GHz is
dominated by either the top of footpoints of a low-lying bundle of
loop-like structures rooted in strong magnetic fields of sunspots.
Such microwave sources are due to gyroresonance emission at the
fourth or even third harmonic of the gyrofrequency, i.e., the
magnetic field strength in the corona reaches 1500--2000~G at a
place where a NLS resides (\authorcite{Uralov2006}
\yearcite{Uralov2006,Uralov2008}; \cite{Nita2011}). Thus,
existence of a NLS in AR~10930 indicates very strong magnetic
field in the corona. Such sources are only observed in active
regions which produce GOES X class flares.

The NLS had a brightness temperature of 0.3--0.5 MK. It was
located above the neutral line, approximately in the middle
between the main sunspots of opposite polarities, where the
eruptive feature EF1 originated. On December 12, the NLS shifted
to the larger northern sunspot of S-polarity and persisted there
until the flare onset (and reappeared after the flare). The flare
started close to the position of the NLS or exactly there. This
situation appears to be a typical one \citep{Uralov2008}.

 \subsubsection{Flare Rise}
When the flare started, the major microwave emission was
contributed by gyrosynchrotron from high-energy electrons. Before
considering the flare course in microwave images we recall that
the turnover frequency of the microwave spectrum in
Figure~\ref{fig:timeprof}e reached 47 GHz during the first peak.
Thus, the microwave source at that time was certainly optically
thick at both operating frequencies of NoRH 17 and 34 GHz. We
therefore limit our analysis with consideration of 17 GHz emission
only and take advantage of polarization data available at 17 GHz.

To produce images at 17 GHz, we used the Fujiki program for the
rise phase of the flare. Very bright microwave sources during the
peaks considerably exceeded 100~MK that makes impossible usage of
standard imaging software of NoRH. Therefore, after 02:24 we used
the program developed by H.~Koshiishi for imaging of extreme
flares. Calibration of the images in brightness temperatures was
performed by referring to total fluxes recorded with NoRP as
proposed by K.~Shibasaki.

\begin{figure*}
  \begin{center}
    \FigureFile(85mm,91mm){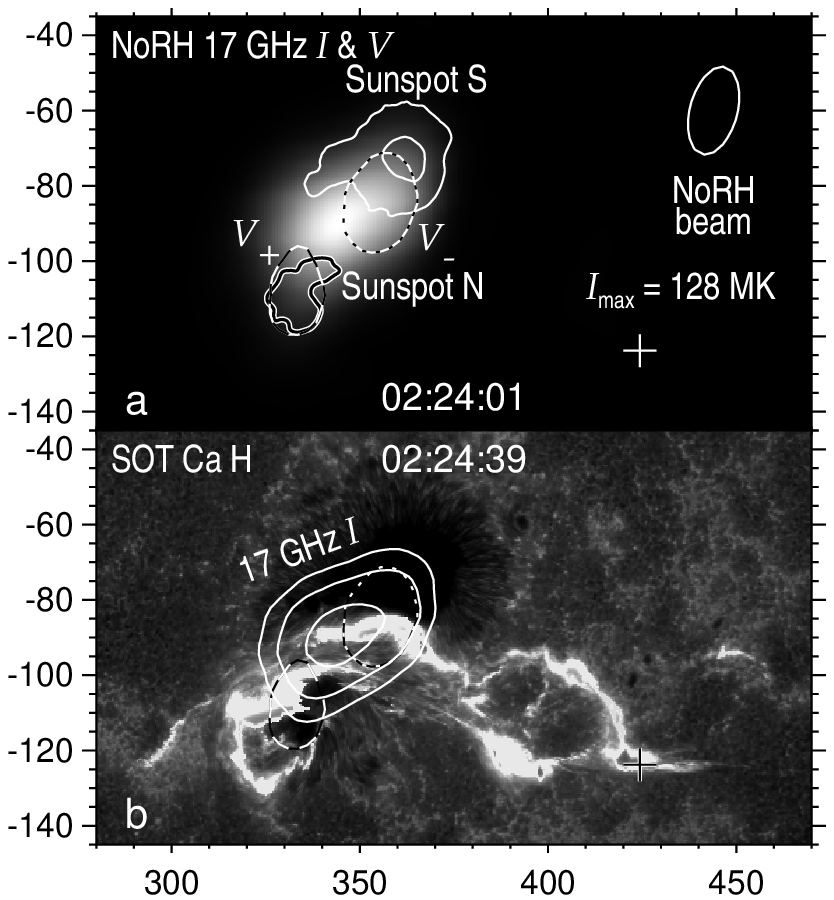}
    \FigureFile(85mm,91mm){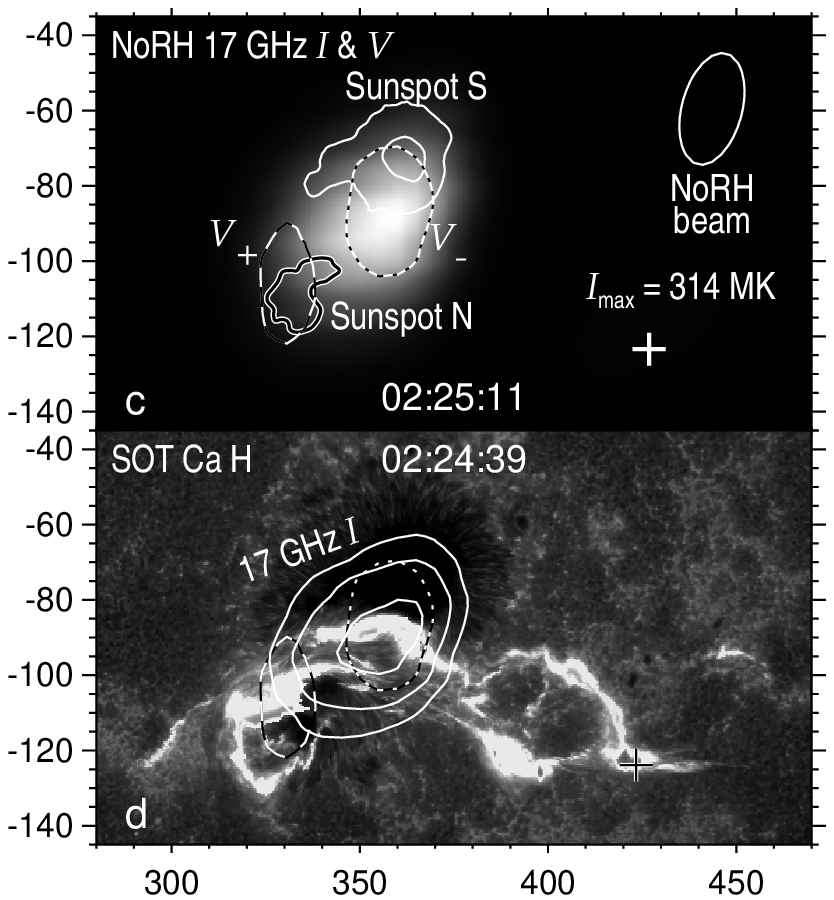}
  \end{center}
  \caption{NoRH and Ca~H (SOT) images during the rise phase (a,b)
and at the first peak (c,d). The contour levels for the 17 GHz
intensity are [25, 50, 100] MK (a) and [62, 125, 250] MK (b). The
contour levels of the polarized component (Stokes $V$) are 50\% of
both positive (dashed) and negative (dotted) maxima. The white
contours in panels (a,c) outline the S-sunspot in the Hinode/SOT
magnetogram, and the black-on-white contours outline the N-sunspot
in the SOHO/MDI magnetogram (the levels are $[-3000, -1500,
+1500]$~G.) The crosses mark the brightness center of the weaker
west microwave source and the corresponding position in the SOT
image.}
  \label{fig:overlay}
 \end{figure*}

One of major difficulties in analyses of microwave images along
with those produced in different spectral domains is their
accurate coalignment, because NoRH does not provide an absolute
pointing. We overcome this problem by comparing the NoRH images
with the flare ribbons shown by Hinode/SOT in the Ca H-line and by
comparing the polarized microwave emission (Stokes $V$ component)
with the magnetograms of Hinode/SOT and SOHO/MDI as shown in
Figure~\ref{fig:overlay}. We have Hinode magnetograms produced
several hours before the event, at 20:30 on December 12, and after
the event, at 04:30 on December 13. The southern N-polarity
sunspot rapidly changed at that time. There is a SOHO/MDI
magnetogram produced at 01:40, close to the onset of the event,
but the large stronger northern S-polarity sunspot, which was
rather stable, is heavily distorted in the magnetogram due to
`high-field saturation'. We combine contours of the stable
negative sunspot taken from the Hinode magnetogram with contours
of the variable positive sunspot from the temporally close MDI
magnetogram, which was not distorted. The two magnetograms were
accurately coaligned with each other and referred to the pointing
of MDI in Figure~\ref{fig:overlay}.

The gray scale background in Figure~\ref{fig:overlay}a presents a
NoRH 17 GHz image in total intensity (Stokes $I$) observed at
02:24, i.e., at the rise phase of the first peak. The solid white
contours outline the levels of $[-3000, -1500]$~G in the
Hinode/SOT magnetogram. The solid black-on-white contour outlines
the $+1500$~G level in the SOHO/MDI magnetogram. The broken
contours correspond to 50\% levels of the polarized emission at 17
GHz (dashed positive, dotted negative). The polarization at 17 GHz
corresponds to the $x$-mode emission with a degree up to $>30\%$
at 02:24. The ellipse in the upper right corner presents a
half-magnitude contour of the NoRH beam. Comparison with the
microwave Stokes $I$ and $V$ data and shows that the source in
total intensity was well resolved and rather large, while the
polarized sources were more compact. The negatively polarized
region was somewhat extended in the east-west direction.

The peak frequency in Figure~\ref{fig:timeprof}e indicates that
the 17 GHz source at that time approached the optically thick
regime. Thus, the major emission in total intensity should be
radiated from upper layers of the source, where the magnetic
fields were weaker (see \cite{DulkMarsh1982,White2003,Kundu2009}).
The positions of the emitting regions in Figure~\ref{fig:overlay}b
correspond to the above considerations. The solid white contours
here outline the levels of $[25, 50, 100]$~MK in total intensity
at 17 GHz, while the maximum brightness temperature over the image
is 128~MK. The broken contours show the same 50\% levels of the
polarization as in the upper panel.

Overall, Figure~\ref{fig:overlay} shows that the total intensity
at 17 GHz was mainly emitted from the broad upper part of the
flare arcade, whose bases are highlighted by the ribbons in the
SOT Ca H-line image. The polarized regions corresponds to the
conjugate legs of the arcade loops. The optical thicknesses of the
polarized regions viewed slightly aside from the arcade top were
most likely less than that of the broad cover source. Microwaves
were mostly emitted by the portion of the arcade between the
sunspots where the magnetic field was much stronger than sidewards
and, probably, the number of high-energy electrons was also
larger.

Nevertheless, microwave images show a suggestion of an additional
weak source moving west from the major flare site during
02:22--02:25. This presumable source overlaps with a region of
strong side lobes from the major source. The reality of this
secondary source is supported by its larger size, gradual shape,
and progressive motion west, all of which are different from the
side lobes. We have roughly measured probable positions of the
brightness centers of the secondary source manually and plotted
them with the white crosses in Figures \ref{fig:overlay} and
\ref{fig:norh_hinode}. The crosses approximately correspond to the
expanding western portions of the ribbons in the SOT and XRT
images. Thus, the weak microwave source moving west displays
development of the flare arcade westward in weaker magnetic fields
apparently caused by the eruptions. On the other hand, closeness
of its measured positions to the developing ribbons indicates that
the coalignment accuracy of all the images is satisfactory. Its
uncertainty presumably does not exceed $5^{\prime \prime}$.

 \subsubsection{First Flare Peak}

Figure~\ref{fig:overlay}c,d presents the same set of images as in
Figure~\ref{fig:overlay}a,b, but the NoRH 17 GHz data correspond
to the first major peak of the flare (02:25:11). As the very high
turnover frequency in Figure~\ref{fig:timeprof}e shows, the
microwave source at that time was certainly dominated by optically
thick emission both at 17 and 34 GHz. The maximum brightness
temperature at 17 GHz reached 314 MK. Nevertheless, the microwave
configuration has not considerably changed. The optically thick
part of the microwave source appears to have broadened as
suggested by a decreased degree of polarization of $< 15\%$
remaining at the upper part of the flare arcade.

The very high turnover frequency of the microwave spectrum at that
time indicates emission from very large number of high-energy
electrons in very strong magnetic fields (see
\cite{DulkMarsh1982,White2003,Grechnev2008protons}). The constancy
of the position of the microwave source while $f_\mathrm{peak}$
drastically increased suggests that the magnetic field strength
remained nearly the same. Thus, the major reason for the change
could be a plentiful ejection of high-energy electrons that should
shift the gyrosynchrotron spectrum right, to considerably higher
frequencies. Note that this flare peak was apparently caused by
the eruption (see Figure~\ref{fig:accelerations}). To find out
what could be a reason for such strong energy release during the
first flare peak, we consider the overall course of the flare
shown by microwave images along with Hinode/SOT and XRT images.

 \subsubsection{Overall Progression of the Flare}

\begin{figure*}
  \begin{center}
    \FigureFile(170mm,115mm){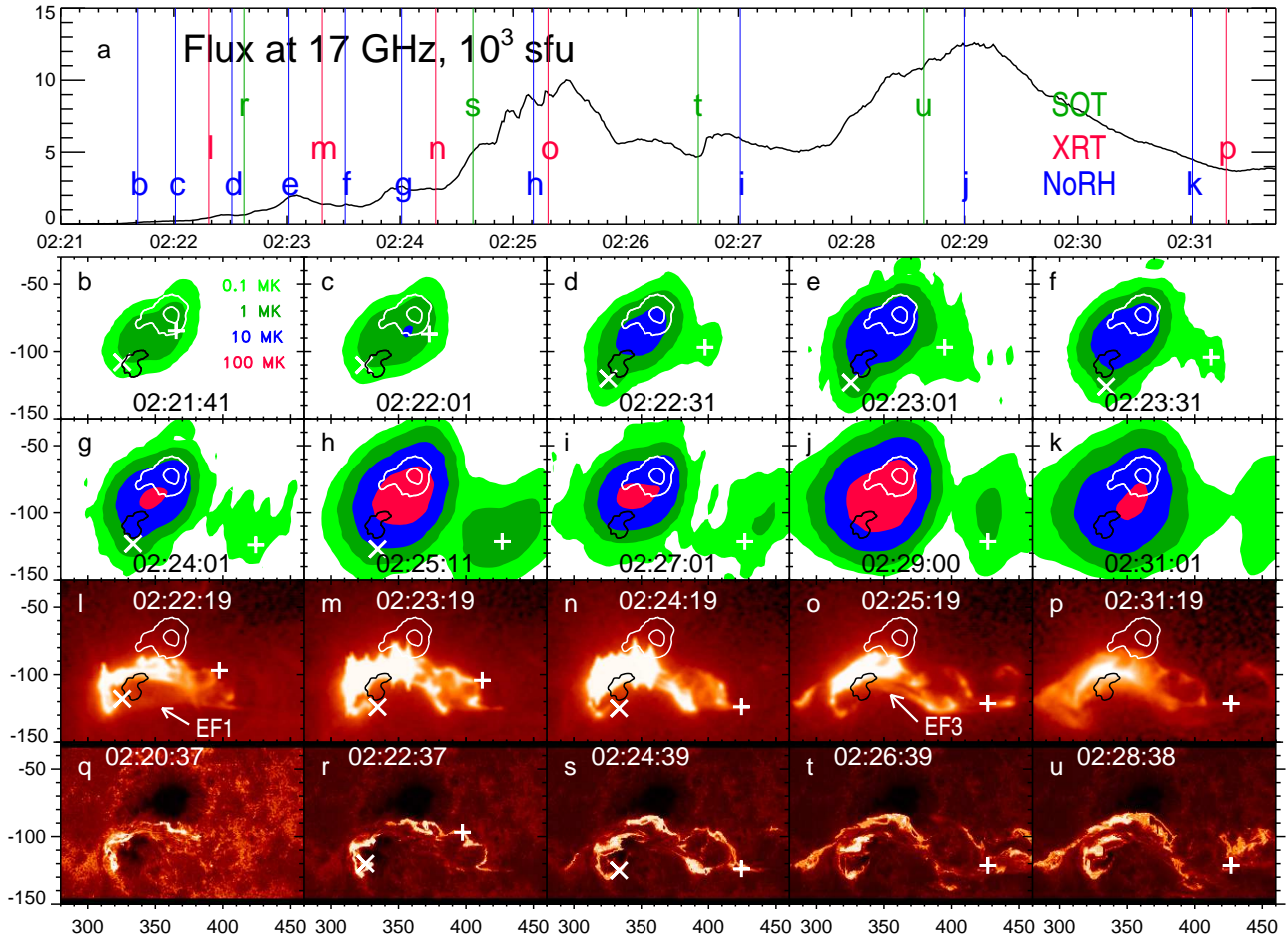}
  \end{center}
  \caption{Overall course of the flare. a)~Total flux time profile at 17 GHz,
b--k)~color contours of NoRH 17 GHz images at
[0.1,~1,~10,~100]~MK, l--p)~Hinode/XRT SXR images,
q--u)~Hinode/SOT Ca H-line images. The contours on top of the NoRH
and XRT images outline the magnetic N-polarity (black, $+1500$~G)
and S-polarity (white, $[-3000, -1500]$~G). The color vertical
lines in panel (a) mark the observing times of the images. The
straight and slanted crosses mark the brightness centers of weaker
microwave sources and corresponding positions in the XRT and SOT
images. Eruptive features EF1 and EF3 are denoted in panels (l)
and (o), respectively; their middle parts are indicated by the
arrows. Eruptive features EF2 and EF4 are beyond the small field
of view presented in the figure.}
 \label{fig:norh_hinode}
 \end{figure*}

The overall course of the flare can be followed from
Figure~\ref{fig:norh_hinode} which shows selected microwave images
(colored shading in two middle rows) along with available XRT
(l--p) and SOT (q--u, lower row) images in comparison with NoRP
time profile at 17 GHz (a, top row). XRT images were produced
every minute, while only one SOT image in the Ca H-line in two
minutes is available. The interval shown in the figure starts from
the rise phase and covers the two major peaks. The colored shading
quantifies the brightness temperatures in the 17 GHz Stokes $I$
images from 0.1 MK to 100 MK (see panel b). The white and black
contours show again the magnetic field strengths (white $[-3000,
-1500]$~G, black $+1500$~G). The straight white crosses present
rough measurements of the weaker 17 GHz source moving westward.
The slanted crosses present analogous measurements of the weak
southeast-to-south extension of the 17 GHz source. The imaging
times are indicated with the vertical lines in the top panel.

The main 17 GHz source persisted between the sunspots being
associated with the upper parts of the arcade loops. Although the
displayed range of brightness temperatures exceeds the nominal
dynamic range of the NoRH, comparison with the XRT and SOT images
confirms suggestions of lower-temperature shading and crosses in
the NoRH images. During the rise phase, the south flare ribbon
developed and moved west, toward the N-sunspot in all the images.
The first major peak occurred when the south ribbon covered the
N-sunspot. Changes in the S-sunspot were not as conspicuous,
because the magnetic flux in this sunspot was much larger than in
the N-sunspot.

The observations confirm strong dependence of the energy release
rate in a flare on the magnetic field strength expected from the
standard flare model. This circumstance was demonstrated by
\authorcite{Asai2002} (\yearcite{Asai2002,Asai2004}). Extreme parameters of
sunspot-associated flares were shown by
\citet{Grechnev2008protons} and \citet{Kundu2009}. Among these
properties are strong hard X-ray and gamma-ray emissions and high
SEP productivity.

After the first major peak, the magnetic flux associated with
strong fields of the N-sunspot mostly reconnected. The microwave
emission did not exceed 150 MK by 02:28. Then the next eruption
caused one more pulse of strong energy release and injection of
accelerated electrons into the flare region. The microwave peak
frequency increased to $\approx$~20 GHz
(Figure~\ref{fig:timeprof}e), and the most 17 GHz source become
optically thick again. The brightness temperature at 02:29 reached
a still higher value of 378 MK possibly due to expansion of the
microwave-emitting region upward into weaker magnetic fields.
Though this peak was higher at 17 GHz than the first one, the
energy release rate and the number of high-energy electrons
probably were less strong.

One of distinctive features of this extreme flare was the increase
of the microwave peak frequency well above 17 GHz and even above
34 GHz during the first peak. Unlike a typical situation, the
microwave-emitting source certainly was not optically thin at 17
GHz during the main peaks. So could be also even with a lower peak
frequency [see \citet{Kundu2009} for detail]. These facts can shed
light on features of the spectral evolution of the flare emissions
in this event (see, e.g., \cite{Ning2008}). The reason for these
extreme properties was involvement in reconnection processes of
strongest magnetic fields rooted in the sunspots in accordance
with the conclusion of \citet{Jing2008}.

\subsection{EUV Shock Signatures}
 \label{s-euv_wave}

\begin{figure}
  \begin{center}
    \FigureFile(85mm,85mm){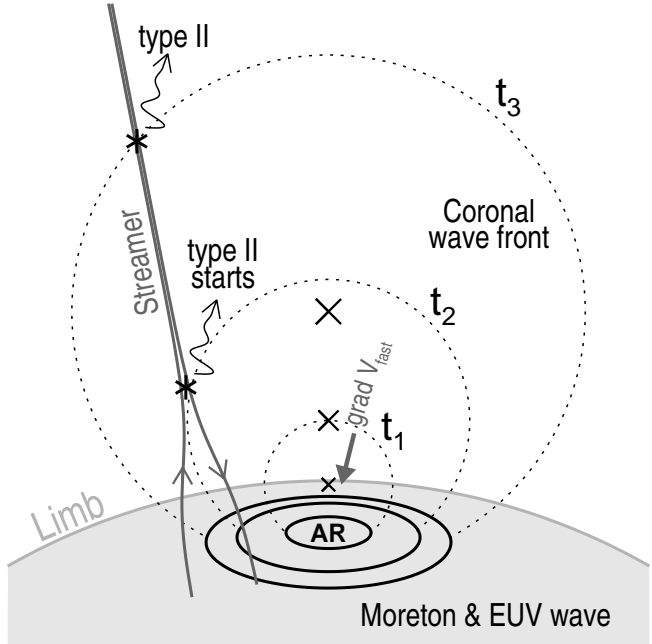}
  \end{center}
  \caption{Fast magnetosonic shock wave excited by an impulsive eruption
in an active region (AR).}
     \label{fig:wave_cartoon}
 \end{figure}

As shown in the preceding sections, at least, two impulsive
eruptions with very strong acceleration of $>15g_{\odot}$
($g_{\odot}$ is the solar gravity acceleration) occurred in the
event. As explained in section~\ref{s-introduction}, coronal shock
waves must have been excited by these impulsive pistons. A cartoon
in Figure~\ref{fig:wave_cartoon} outlines a conception of the
front shape and velocity of a coronal wave excited in an active
region (AR). The positions of the wave front in the corona at
three different times $t_1$, $t_2$, and $t_3$ are presented with
the dotted curves, and their corresponding near-surface traces are
shown with the solid ellipses. The arrow $\mathbf{grad}\,
V_\mathrm{fast}$ represents the conditions in the low corona above
the active region favoring the wave amplification and formation of
a discontinuity at $t_1$. The blast-like wave is expelled from the
AR's magnetosphere into regions of weaker magnetic fields. The
shape of the wave front in the low corona is determined by the
Alfv{\' e}n speed distribution. The front shape is close to an
oval, possibly oblate in its upper part with a moderate intensity
of the wave. The center of the oval progressively displaces
upward, as increasing slanted crosses show. If the shock wave is
strong enough, then the shape of its front should be an oval
expanded in its upper part. Crossing by the shock front of the
current sheet inside a coronal streamer excites type II radio
emission.

Probable signatures of shock waves are large-scale transients in
extreme ultraviolet (EUV) known as `EUV waves' (or `EIT waves').
Such a transient was really present in this event (see, e.g.,
\cite{Asai2008,Liu2008,Li2009,Attrill2010,Nitta2012}). However, no
analysis of this transient was carried out besides a conclusion
about association of this `EUV wave' with a shock.

The routine imaging rate of SOHO/EIT of 12 min was insufficient to
study propagation of such transients in detail. We therefore
combine EIT observations with those of GOES-12/SXI, although its
images reveal `EUV waves' poorer than EIT and heavily suffer from
the scattered light. To understand what the images show, we
calculated an expected propagation of a shock wave over the
spherical solar surface assuming homogeneous corona in the way
described in section~\ref{s-techniques} and compared the
calculated positions of the front with its presumable traces in
real images. We used as input parameters probable excitation times
of shocks corresponding to the peaks of acceleration of EF2 and
EF3 and one of well-defined wave fronts. Then we attempted to find
other signatures of the shock propagation and corrected the input
parameters to fit them better.

\begin{figure*}
  \begin{center}
    \FigureFile(145mm,139mm){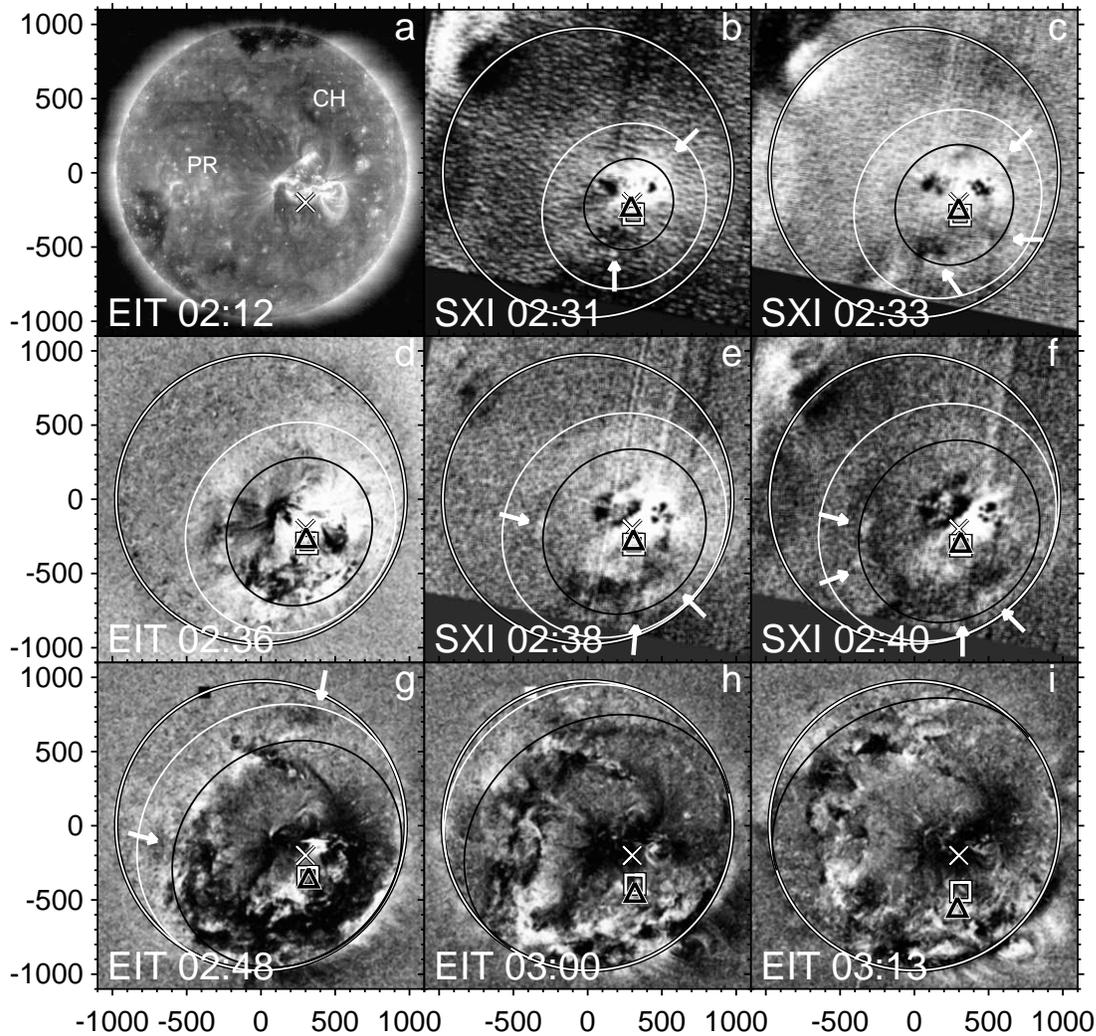}
  \end{center}
  \caption{EUV traces of two shock waves in SOHO/EIT and GOES-12/SXI
images. a)~Pre-event EIT 195~\AA\ image. `CH' is a coronal hole,
`PR' is a plage region. The slanted cross shows in all images the
initial position of the wave center. b--i)~Running-difference
ratios of EIT 195~\AA\ (d, g, h, i) and SXI (b, c, e, f) images.
The ellipses present the calculated intersections of the
spheroidal shock fronts with the spherical solar surface (white
shock~1, black shock~2). The arrows indicate suggestions of shock
traces. The white squares mark the current epicenters of shock~1.
The black triangles mark the current epicenters of shock~2.}
 \label{fig:eit_sxi}
\end{figure*}

Attempts to fit in this way the observed shock signatures as
propagation of a single front have resulted in discrepancies with
the observations. The calculations have been reconciled with the
real images when we considered two shock fronts following each
other. One shock (shock~1) was excited by eruption EF2 at 02:23,
and the second one (shock~2) was excited by eruption EF3 at 02:27.
The results are shown in Figure~\ref{fig:eit_sxi}.

Figure~\ref{fig:eit_sxi}a shows an EIT 195~\AA\ pre-event image
following by a set of EIT and SXI running-difference ratios in
panels (b--i). The initial position of the wave center is marked
by the slanted cross. The pre-event image reveals some
inhomogeneities in the corona that can affect propagation of a
shock wave. One of them is a darker region CH northwest from the
active region resembling a coronal hole. A SOHO/MDI magnetogram
shows in this region an enhanced, predominantly unipolar
(negative) magnetic field. Thus, the Alfv{\' e}n speed was
enhanced above this region. Other inhomogeneities are plage
regions PR east from AR~10930 and a coronal hole farther eastward.

The white and black ellipses present intersections of two
spheroidal wave fronts with the spherical solar surface calculated
for the wave propagation in a homogeneous medium. This simplified
approximation is only convenient for a portion of the wave front
running along the homogeneous solar surface. The ellipses were
calculated by referring to obvious wave traces in a few images. On
the other hand, the ellipses hint at shock suggestions in other
images, some of which are indicated by the arrows. For example,
the northwest brightening marked by the arrows in
Figure~\ref{fig:eit_sxi}b,c might be due to the scattered light;
however, its expansion corresponds to the expected propagation of
the second shock front. Also, the dark patches in these
running-difference ratios (Figure~\ref{fig:eit_sxi}b,c) preceded
by faint brightenings might be due to a propagating disturbance.
Thus, if some of the mentioned features visible in the GOES/SXI
images are real, then their positions correspond to expected
traces of the shocks.

The two fronts are especially pronounced in
Figure~\ref{fig:eit_sxi}d. The compact east brightenings indicated
by the arrows in Figure~\ref{fig:eit_sxi}e,f might be due to pass
of the shock front over the west plage region. The north dark
patch just behind the white ellipse indicated by the arrow in
Figure~\ref{fig:eit_sxi}g certifies the pass of a disturbance
there. Similarly, traces of the disturbance are visible behind the
calculated front of the first shock in
Figure~\ref{fig:eit_sxi}h,i.

The actual fronts ran faster than the calculated ellipses in the
northwest region CH with a higher Alfv{\' e}n speed. This
deviation is expected for a shock front. The plage regions PR and
a coronal hole eastward also affected propagation of the shocks.
Remarkable is a progressive displacement of the wave epicenters
(the squares and triangles) toward the southern polar coronal
hole. We had to introduce this displacement to co-ordinate the
calculated ellipses with actual large-scale shapes of the fronts.
The progressive shift is a property of 3D fast-mode MHD shock
waves whose propagation is determined by the Alfv{\' e}n speed
distribution (Figure~\ref{fig:wave_cartoon};
\authorcite{Grechnev2011_I}
\yearcite{Grechnev2011_I,Grechnev2011_III};
\cite{AfanasyevUralov2011}).

To our knowledge, this is the first case of two shocks following
each other along the solar surface with a difference between their
excitation times as small as four minutes revealed from imaging
observations. Detection of the two separate shock fronts has
become possible presumably because the first shock was
considerably faster than the trailing one.

\begin{figure}
  \begin{center}
    \FigureFile(85mm,85mm){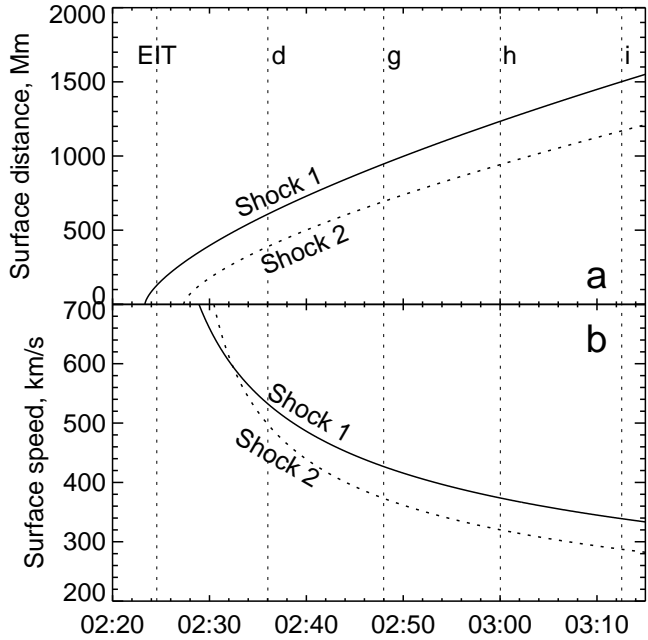}
  \end{center}
  \caption{Distance-time (a) and velocity-time (b)
plots of the two shocks following each other along the solar
surface. The distances are measured from the initial wave center
(the slanted cross in Figure~\ref{fig:eit_sxi}). The onset times
are 02:23:20 for shock~1 and 02:27:20 for shock~2; $\delta = 2.0$
for both shocks. The vertical dotted lines mark the imaging times
of EIT, and the labels denote the corresponding panels in
Figure~\ref{fig:eit_sxi}.}
     \label{fig:euv_wave_plots}
 \end{figure}

Figure~\ref{fig:euv_wave_plots} presents kinematical plots for
propagation of the two shock fronts along the spherical solar
surface corresponding to the ellipses in Figure~\ref{fig:eit_sxi}.
To facilitate comparison of the plots with the images, the
vertical dashed lines mark the times of the EIT images shown in
Figure~\ref{fig:eit_sxi}. The labels `d, g, h, i' in
Figure~\ref{fig:euv_wave_plots}a indicate the corresponding panels
in Figure~\ref{fig:eit_sxi}. Both shock fronts monotonically
decelerated. \citet{Asai2008} estimated the speed of the ``EIT
wave'' to be `\textit{$570 \pm 150$~km~s$^{-1}$ in the southeast
direction}' probably relating the estimate to 02:36. We remind
that the shock fronts in Figure~\ref{fig:eit_sxi} were calculated
for isotropic shock propagation along the surface, and therefore
the plots in Figure~\ref{fig:euv_wave_plots} represent an
azimuthally-averaged kinematics of the wave fronts.

The average propagation velocities of the first and second `EUV
wave's fronts at a distance of $1R_{\odot}$ from their epicenters
in the active region were about 500 and 380 km~s$^{-1}$,
respectively. Typical propagation velocities of `EUV waves' at
such distances are close to the fast-mode speed in the low corona
above the quiet Sun and, most likely, rarely exceed 300
km~s$^{-1}$ \citep{Mann2003,Grechnev2011_III}. Such high
propagation velocities in the 2006 December 13 event evidence
nonlinear character of the near-surface magnetosonic wave and a
high probability of its shock-wave regime. However, the supersonic
propagation velocity of the visible wave disturbance does not
guarantee that the observed plasma compression moves together with
the shock front. The top of a nonlinear wave moves faster than its
foot even before the formation of the discontinuity. To reveal a
velocity jump evidencing a shock front, one should directly
measure the kinematics of magnetic structures (e.g., coronal
loops) after the pass through them of the near-surface ``EUV
wave''. The observational data do not allow such direct
measurements. To get further support to the shock-wave regime of
the wave front, we consider in the next section its portion
responsible for the appearance of the type II emission, which is
believed to be due to a shock front.

\subsection{Dynamic Radio Spectrum}
 \label{s-dynamic_spectrum}

As mentioned, radio emission caused by the 2006 December 13 event
was extraordinarily strong at meters and especially decimeters,
with total fluxes up to $0.5 \times 10^6$ sfu
(Figure~\ref{fig:timeprof}g). The huge decimetric burst was
apparently due to a coherent mechanism, probably ECM
\citep{Kintner2009}. This coherent emission complicates
consideration of weaker type II and type IV bursts. The long-wave
part of the type II burst was discussed by
\authorcite{Firoz2011} (\yearcite{Firoz2011,Firoz2012}) and
analyzed up to kilometers by \citet{Liu2008}.

\begin{figure}
  \begin{center}
    \FigureFile(85mm,85mm){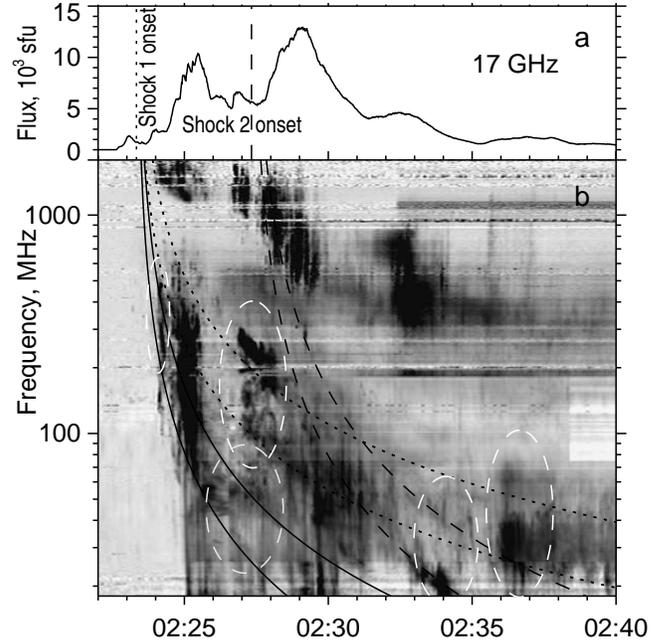}
  \end{center}
  \caption{Dynamic radio spectrum composed from Culgoora, Learmonth, and
Callisto/SSRT data (b) in comparison with the microwave time
profile at 17 GHz (NoRP). Presumable onset times of shock~1
(02:23:20) and shock~2 (02:27:20) are denoted with the vertical
broken lines in top panel. Some type II bands detectable in the
spectrum are outlined with the black curves calculated for
expected shock propagation. The solid and dotted pairs of curves
outline signatures of shock~1, and the dashed one outlines
shock~2. The dashed white ovals indicate most pronounced harmonic
pairs of type II bands.}
 \label{fig:dyn_spectrum}
 \end{figure}

The short-wave portion of the type II and type IV bursts is most
important for our analysis. To cover the whole frequency range of
interest and make the dynamic spectrum clearer, we have combined
the data of Culgoora, Learmonth, and Callisto/SSRT spectrometers.
The composite spectrum is presented in
Figure~\ref{fig:dyn_spectrum}b in comparison with the microwave
total flux at 17 GHz in Figure~\ref{fig:dyn_spectrum}a.

The dynamic spectrum shows probable manifestations of type III,
II, and IV bursts. A few strong trains of coherent emission
(presumably ECM) correspond to the microwave peaks in
Figure~\ref{fig:dyn_spectrum}a. A drifting type IV burst masked by
stronger coherent bursts suggests emission from electrons trapped
in an expanding flux rope of the developing CME. A detailed
analysis of the type IV emission requires involvement of
longer-wave data and is beyond our scope.

The structure of the type II burst is complex suggesting emissions
from different coronal streamers stressed by two shock fronts (see
Figure~\ref{fig:wave_cartoon} and \cite{Grechnev2011_I} for
details). Some well-pronounced pairs of type II bands are
indicated in the figure with the white dashed ovals. To identify
these manifestations with a particular shock and follow the drift
rate for each of them, we used the power-law approximation of
shock wave propagation (see section~\ref{s-techniques} and
\authorcite{Grechnev2011_I}
\yearcite{Grechnev2011_I,Grechnev2011_III}). To recognize
signatures of the shocks in the dynamic spectrum, we firstly
analyzed the spectra of Culgoora, Learmonth, Callisto/SSRT, and
HiRAS separately, and then verified the results by comparing them
with each spectrum.

The results, which we have reached so far, are presented in
Figure~\ref{fig:dyn_spectrum}b with the black pairs of curves
(fundamental \& second harmonic). The solid and dotted pairs of
curves outline signatures of shock~1 whose presumable onset time
of 02:23:20 is denoted with the dotted line in
Figure~\ref{fig:dyn_spectrum}a. These two pairs of bands were
presumably generated by two parts of the shock~1 front passing in
two different streamers. Similarly, all the dashed lines are
related to shock~2 with an onset time of 02:27:20. The onset times
have been actually estimated in attempts to achieve a best
correspondence of the outlining curves with the actual signatures
of the shocks in the dynamic spectrum and used afterwards in
outlining shock traces in Figure~\ref{fig:eit_sxi}. The
kinematical curves of the near-surface ``EUV waves'' in
Figure~\ref{fig:euv_wave_plots}, on the one hand, and the
outlining curves corresponding to the coronal shocks in
Figure~\ref{fig:dyn_spectrum}, on the other hand, were obtained in
self-consistent calculations to fit both EUV-imaging and radio
data. This fact confirms that both coronal waves were shock waves,
at least, as early as the corresponding type II bursts started.
Their onset times were about 02:24 for the first-wave bands and
02:33 for the second-wave ones.

The outlining curves presented in Figure~\ref{fig:dyn_spectrum}b
were calculated for shock waves, which were impulsively generated
and freely propagated afterwards. While the shocks were most
likely excited by the impulsive-piston mechanism, their further
behavior resembles decelerating blast waves.

\subsection{CME}
 \label{s-cme}

\begin{figure*}
  \begin{center}
    \FigureFile(170mm,110mm){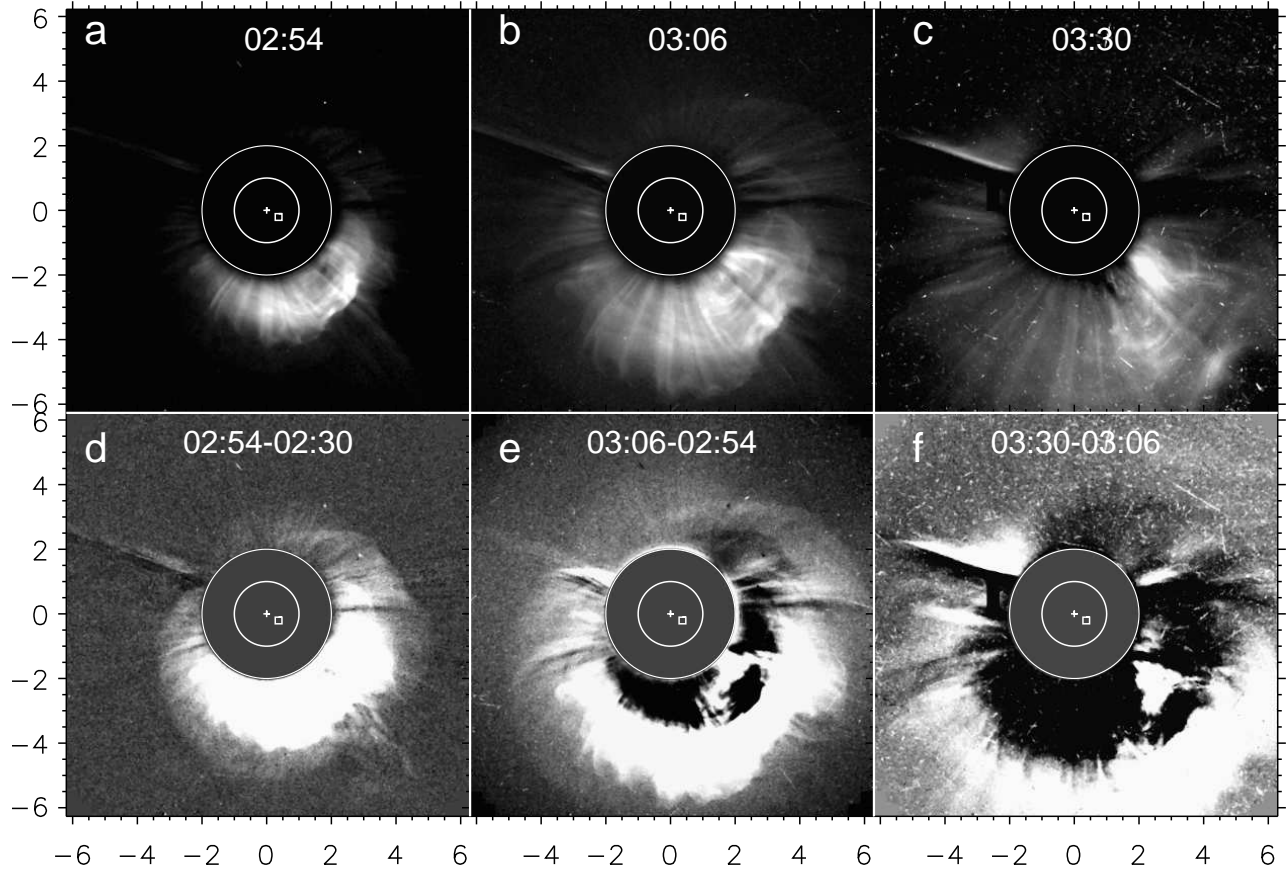}
  \end{center}
  \caption{LASCO/C2 images of the CME developed in the 2006 December
13 event. Top row (a--c): fixed-base ratios present the structure
of the CME. Bottom row (d--f): enhanced-contrast running
differences reveal faint features associated with a leading edge
presumable formed by the shock front. Apparent inner features in
the running-difference images are due to subtraction of preceding
images and can be deceptive. The inner circle denotes the solar
limb, and the outer one denotes the internal boundary of the C2
field of view ($2R_{\odot}$). The cross marks the solar disk
center. The square marks AR~10930. The axes show the distances
from the solar disk center in solar radii.}
    \label{fig:lasco}
     \end{figure*}

The 2006 December 13 event has produced a fast decelerating halo
CME with an estimated average speed of 1774~km~s$^{-1}$ according
to the SOHO LASCO CME Catalog
(http:/\negthinspace/cdaw.gsfc.nasa.gov/CME\_list,
\cite{Yashiro2004}). The CME is shown in Figure~\ref{fig:lasco} in
two representations approximately corresponding to their
appearance in the Catalog. The top row (a--c) contains fixed-ratio
images, which allow one to analyze the structure of the CME. The
bottom row (d--f) contains enhanced-contrast running differences
produced by subtraction from each image of the immediately
preceding one. Such images reveal weakest manifestations of the
leading edge of a CME and possibly ahead. However, the appearance
of a CME in such images can be different from its real structure,
because subtraction of a preceding image and heavy enhancement of
the contrast produce deceptive effects behind the leading edge
\citep{ChertokGrechnev2005,Bogachev2009}.

Figures~\ref{fig:lasco}a,d show a weak suggestion of the streamer
at a position angle PA $\approx 225^{\circ}$, above active region
10930 from which the CME originated (this streamer is distinct in
images presented in the CME catalog). The images in the top row
reveal a complex structure of the transient. The outer radial
features could be partly due to deflected coronal rays. Some of
them, especially around PA of $180^{\circ}$, look like loops which
suggests that they were probably constituted by an expanding
arcade. Transversal structures are visible inside the presumable
arcade. The brightest part of the CME extended approximately from
AR~10930 along the streamer being probably a core flux rope.
Keeping in mind that the CME was formed from four sequential
eruptions and most likely contained an expanding pre-eruption
arcade, one might find some associated features in the CME, while
we do not try to establish their on-to-one correspondence.
Overall, expectations from the eruptions in AR~10930 appear to
correspond to the CME structural components and their
orientations.

Figures~\ref{fig:lasco}d--e reveal indications of a shock wave
propagating ahead of the CME body. The images in
Figures~\ref{fig:lasco}d,e resemble an umbrella blowed by strong
wind from inside. The shock conspicuously deflected large
streamers in Figures~\ref{fig:lasco}e,f, while the outer trace of
its front is outlined by the faint halo edge of the transient. A
portion of the flux rope core with a central PA $\approx
225^{\circ}$ is visible in these two images.

\begin{figure}
  \begin{center}
    \FigureFile(85mm,51mm){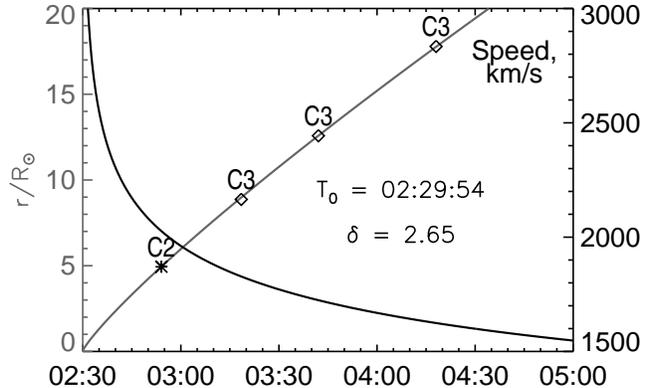}
  \end{center}
  \caption{Height-time (gray, left axis) and velocity-time
(black, right axis) plots of the CME's leading edge fitted as a
shock wave. The symbols present measurements in the SOHO LASCO CME
Catalog.}
     \label{fig:lasco_fit}
 \end{figure}

Measurements in the CME catalog refer to the fastest feature of an
observed transient. The measurements in the catalog for this CME
carried out at a position angle of $\approx 193^{\circ}$ are most
likely related to the shock. Figure~\ref{fig:lasco_fit} presents
the measurements from the catalog (symbols and left axis) along
with the gray fit calculated for shock wave propagation. The
descending black curve and the right axis present the
corresponding velocity-time plot.

The optimizing software has found the density falloff exponent of
$\delta = 2.65$ corresponding to the mid-latitude Saito model
\citep{Saito1970} and the shock onset time of about 02:30 (at the
position of AR~10930). This result is consistent with an
expectation that after some time, the upwards-propagating sections
of the two shocks following each other (shock~1 excited by
eruption EF2 and shock~2 excited by eruption EF3) should merge
into a single faster shock with a seemingly later onset time
\citep{Grechnev2011_I}. This effect is schematically similar to
the situation where two pistons located close to each other are
substituted with a single piston, which sharply increases its
expansion speed twice. After a transition process, a single shock
would remain instead of two. This circumstance also hints at a
possibility of the earliest shock produced by eruption of EF1 at
about 02:21, which was probably reached by the shock produced by
eruption EF2 and merged into a single strong shock~1. This
possibility is in accordance with the assumption of
\citet{Asai2008} that their blue-shifted feature BS2 observed at
02:23:14 was related to a shock.

Comparison of Figures \ref{fig:lasco}b and \ref{fig:lasco}e shows
that the heavy image processing has exaggerated the weight of the
arcade-like component so that the visible central position angle
of the transient looks more like $\approx 190^{\circ}$. This
illusive offset of the CME relative to the position angle of the
parent active region was a subject of concern of
\citet{Sterling2011} and forced the authors to conclude that the
major eruption occurred away from strong magnetic fields. However,
non-subtracted images in Figure~\ref{fig:lasco}a--c show that the
major flux rope of the CME expanded nearly radially above the
active region, along the corresponding streamer. The eruptions and
flare occurred in strongest magnetic fields, and the course of the
flare appears to be well described by the standard model.

\section{Discussion}
 \label{s-discussion}

Preparation of the extreme 2006 December 13 event was indicated by
the long-lived microwave neutral line source observed for a week
before. The extreme properties of the event were determined by the
fact that the eruptions and flare occurred in strongest magnetic
fields above the sunspot umbrae. The umbra of the smaller southern
sunspot was entirely covered by the flare ribbon. The other ribbon
noticeably intruded into the umbra of the larger northern sunspot.
These facts and properties of the flare indicate involvement in
reconnection of large magnetic flux and agree with the conclusion
of \citet{Jing2008} that `\textit{high energy release regions tend
to be concentrated in local strong field regions}'.

Previously extreme properties of sunspot-associated flares were
stated by \citet{Grechnev2008protons} and \citet{Kundu2009}.
According to their conclusions, strong emissions in hard X-rays
and gamma-rays are also expected in such events. There is no
corresponding information for the 2006 December 13 event. To our
knowledge, the only detector of hard electromagnetic emissions in
2006 was RHESSI, but it did not observe the strongest first flare
peak due to nighttime (see Figure~\ref{fig:timeprof}f).

Analysis of various aspects of the 2006 December 13 event sheds
light on some long-standing issues. One of them is related to a
deviation from the Neupert effect.

\subsection{The Neupert Effect}
 \label{s-neupert_effect}

From comparison of 1-min GOES data with a total flux time profile
at 15.4 GHz \citet{StruminskyZimovets2008} came to a conclusion
drawn about a deviation of emission in this event from the Neupert
effect \citep{Neupert1968}. To investigate into this issue, we
used 3-sec GOES data. Their discrete `staircase' character
disfavoring differentiation was overcome by means of smooth cubic
spline approximation with cross-validation for estimation of the
smoothing parameter. The result presented in
Figure~\ref{fig:timeprof}b demonstrates that the derivative of the
0.5--4~\AA\ channel responded to the first (02:25:11) and third
(02:32:30) peaks, while the second peak at 02:29:00 was absent.
The derivative of the 1--8~\AA\ channel is inconclusive, because a
peak indicated with the question mark is most likely an artifact.
Indeed, there was a deviation from the Neupert effect which
\citet{StruminskyZimovets2008} explained by `\textit{an effective
escape of accelerated particles into interplanetary space rather
than their precipitation into dense layers of the solar
atmosphere}'. However, if accelerated electrons escaped, then no
corresponding microwave peak occurred, and the Neupert effect
worked perfectly in such a case. Strong flux of precipitating
electrons is evidenced by the strong hard X-ray burst at that time
in Figure~\ref{fig:timeprof}f.

A reason for the deviation from the Neupert effect is apparently
due to deficiency of the soft X-ray emission. This emission is
produced by evaporated plasmas confined in closed coronal
structures. The next eruption most likely opened some of them thus
releasing confined thermal plasma. Expansion of escaping plasma
should dramatically reduce the emission measure in soft X-rays and
diminish the second peak in the derivative. Thus, a deviation from
the Neupert effect in a multi-peak event might be indicative of an
additional eruption. Furthermore, the presence of two or more
major peaks in the hard X-ray or microwave time profile of a flare
might be indicative of more than one eruption. This conjecture is
supported by the fact that more than one type II burst are
registered in some events.

\subsection{Shock Waves}
 \label{s-shock_waves}

Accordingly, at least, two (or possibly even three) shock waves
developed in the 2006 December 13 event. The onset times of
shock~1 and shock~2 correspond to the acceleration peaks of two
eruptive features EF2 and EF3, respectively. Both shocks developed
about 2 min before the corresponding flare peaks. The initial
positions of the shock wave centers confirm their association with
the eruptions. Being excited by the impulsive-piston mechanism,
the shock waves detached from the pistons and quasi-freely
propagated for some time like decelerating blast waves.
Manifestations of the two shocks following each other in EUV
images and in the dynamic radio spectrum correspond to each other.
Local deviations of the wave fronts in regions of increased
fast-mode speed from isotropic propagation and progressive
displacement of the wave centers toward the coronal hole confirm
their MHD wave nature. All of these circumstances strongly support
the expectations of \citet{Grechnev2011_I} for the development and
evolution of shock waves.

\citet{Attrill2010} studied evolution of coronal dimmings starting
from their appearance and especially the post-CME recovery in two
events, one of which was 2006 December 13. The authors considered
a possibility for remote dimmings (which they called secondary) to
be formed due to the pass of the shock front. This mechanism was
assessed to be insufficient to account for long-lived dimmings,
because consequences of a shock wave were considered to be
reversible at short time scales like those of weak disturbances.

However, Figure~\ref{fig:eit_sxi} clearly shows that expansion of
the zone of dimming was directly associated with propagation of
the coronal shock wave over the solar surface. The wave front is a
moving boundary of a large-scale MHD flow, which accompanies the
expansion and eruption of magnetic structures associated with a
CME drawing away from the Sun. At the stage, when the CME is
formed and accelerates, this flow has a character of a lateral
expansion of coronal magnetoplasmas being similar to a blow of
wind directed outward. Consequences of this flow for a coronal
loop depend on its size and orientation. The lower portion of the
wave front is tilted toward the solar surface
\citep{Uchida1968,AfanasyevUralov2011}, and therefore low loops
and filaments should be initially pressed sideward-down, thus
producing brightenings just behind the wave front. A very high
coronal loop should be initially displaced sideward-up like a
sail. Stretch of such loops should cause plasma outflow from their
basis, thus producing dimmings. The recovery of the dimmings
formed in such a way requires a considerable time, much longer
than the timescale of their appearance. Thus, the development of
remote dimmings also could be due to the pass of the shock front.

EIT and SXI images of this event have provided a unique
opportunity to reveal two shocks following each other along the
solar surface. On the other hand, the two shock fronts propagating
upwards most likely merged into a single stronger shock, whose
propagation is consistent with the observed expansion of the CME's
leading edge. The shock monotonically decelerated. Deceleration of
a shock front along with a nearly constant established speed of a
CME behind it suggests that their extrapolated height-time plots
should intersect after some time. The actual outcome depends on
the CME speed. If the CME is slow as was the case in the event
addressed by \citet{Grechnev2011_III}, then the shock should
eventually decay into a weak disturbance. If the CME is fast which
was the case in the 2006 December 13 event, then the
blast-wave-like shock should eventually transform into a bow shock
ahead of the CME. The latter scenario is confirmed by velocity
profiles of shocks ahead of interplanetary CMEs (ICMEs) measured
in situ at large distances from the Sun. These velocity profiles
are typical of bow shocks continuously pushed by trailing pistons.
This was also the case for the interplanetary shock wave developed
in the 2006 December 13 event analyzed in detail by
\citet{Liu2008}, who followed its subsequent propagation up to
2.73 AU.

Excitation of shock waves and their evolution turns out more
complex than traditionally assumed. Presumption of a `CME-driven
shock' directly excited exceptionally in the bow-shock scenario by
the outer CME surface at a considerable height well after the
flare appears to be misleading oversimplification. Case studies of
very different events with importance from less then GOES C-class
(see \cite{Grechnev2011_I}) up to the extreme X-class event in the
present study confirm the same impulsive-piston shock formation
inside a developing CME. Actually sharp MHD disturbances, which
transform into shock waves, appear at the rise phase of hard X-ray
and microwave burst being ready to accelerate particles to high
energies and only dampen and decelerate afterwards. Thus,
considerations of relative timing of energetic particles with
respect to a flare do not provide any support to their
acceleration by shock waves.

Prompt acceleration of protons to high energies simultaneously
with electrons during flares is confirmed in occasional
observations of the $\pi^0$-decay emission, which is generated by
$>300$ MeV protons precipitating into dense layers of the solar
atmosphere
\citep{Grechnev2008protons,VilmerMacKinnonHurford2011,Kurt2013}.
On the other hand, escape of flare-accelerated particles from an
active region should be favored by stretch of closed magnetic
configurations in the course of CME lift-off (K.-L.~Klein, 2011,
private communication). If the expanding magnetic flux rope of the
CME reconnects with a coronal streamer (e.g., above the parent
active region), then the particles trapped in the flux rope gain
access to magnetic fields open into the interplanetary space (see
also \cite{Aschwanden2012}). For this reason, the presence of an
apparent delay of particle release near the Sun after the flare
seems to favor particle acceleration in the flare rather than by a
shock.

\subsection{Comments on Comparison of GLE69 and GLE70}

As mentioned, the 70-th ground level enhancement (GLE) of cosmic
ray intensity produced by the 2006 December 13 event was analyzed
in several studies (e.g.,
\cite{Reames2009,Li2009,Aschwanden2012,Nitta2012}). The
conclusions about a possible solar source of GLE particles are
different. Some of them allow contributions from both
flare-related and shock-related acceleration, and some others
favor the only certain source. A number of studies endeavors to
reach certain conclusions in statistical or comparative analyses.

Such a comparative analysis has been undertaken by
\authorcite{Firoz2011} (\yearcite{Firoz2011,Firoz2012}) who
compared the 2006 December 13 event responsible for GLE70 with the
2005 January 20 event responsible for GLE69. The authors'
conclusions are nearly opposite for these two events, while our
results as well as those of \citet{Grechnev2008protons} for the
GLE69 event appear to be very similar. In particular, flares in
both events occurred in very strong magnetic fields above the
umbrae of sunspots and produced very strong microwave bursts with
peak frequencies exceeding 25~GHz. The major circumstances which
have led
\authorcite{Firoz2011} (\yearcite{Firoz2011,Firoz2012}) to
contrasting these two events are as follows.

\begin{enumerate}

\item
 Hard X-ray and gamma-ray emissions were considerably stronger
in the event responsible for GLE69 with respect to the GLE70
event.

-- However, there is no information about these emissions in the
strongest peak on 2006 December 13 because of RHESSI nighttime
(Figure~\ref{fig:timeprof}f).

\item
 Relative timing of particle injection and flare emissions disfavor
their association.

-- Such considerations are inconsistent due to the preceding item
and conclusions in section~\ref{s-shock_waves}.

\item
 The GLE69-related CME was considerably slower (882 km~s$^{-1}$) than the
GLE70-related one (1773 km~s$^{-1}$).

-- Indeed, the SOHO LASCO CME Catalog presents the speed of 882
km~s$^{-1}$ for the 2005 January 20 event, but with a note that it
might be underestimated due to strong contamination of LASCO
images by energetic particles. The note refers to an estimate of
\citet{Gopalswamy2005} of 3242 km~s$^{-1}$. Different estimates
for the speed of this CME presented by \citet{Grechnev2008protons}
range from 2000 to 2600 km~s$^{-1}$. Thus, the situation was
opposite to the assumption of
\authorcite{Firoz2011} (\yearcite{Firoz2011,Firoz2012}): the
GLE69-related CME was considerably faster than the GLE70-related
one.

\item
 The type II burst (in a range of 0.1--1 MHz) in the GLE69 event was
`less dynamic' than that in the GLE70 event.

-- Such a comparison should necessarily refer to locations of the
emission sources. \citet{Grechnev2011_I} have confirmed the idea
of \citet{Uralova1994} that a type II emission appears in a
flare-like process running along the current sheet of a coronal
streamer stressed by a shock front. The images of the 2005 January
20 event in the CME Catalog show plasma outflow in both western
streamers closest to the eruption site. These streamers probably
could not generate the type II burst. Its source region could be
in a remote eastern streamer. Before reaching it, the flank of the
shock wave should have been considerably decelerated. On the other
hand, no obstacles are seen for the appearance of a type II burst
from the closest western streamers in the 2006 December 13 event.

\end{enumerate}

There is no convincing reason for contrasting the events
responsible for GLE69 and GLE70. The solar events were rather
similar in their major properties which determined their
extremeness, while several qualitative and quantitative
differences were certainly present.

\section{Conclusion}
 \label{s-conclusion}

Our multi-spectral analysis of the extreme 2006 December 13 event
involving microwave total flux measurements and imaging data has
revealed its important properties. The observations along with
their quantitative descriptions constitute a consistent picture of
the event that clearly shows the following.

\begin{enumerate}

\item
 Development of this eruptive flare appears to be well described
by the standard model. The flare arcade developed in the course of
a few eruptions.

\item
 The flare episodes were caused by the eruptions being delayed
after them. The repetitive eruptions additionally opened the
coronal configuration permitting escape of evaporated plasmas that
has resulted in a deviation from the Neupert effect.

\item
 The flare emissions were strongest and hardest when flaring occurred in
the strongest magnetic fields above the sunspot umbrae.

\item
 At least, two shock waves were excited by the eruptions as impulsive
pistons inside a developing CME and before the related flare
peaks. Then the shock waves quasi-freely propagated like
decelerating blast waves.

\item
 The two shock waves propagating upward most likely merged into a
single stronger shock, which constituted the outer halo envelope
of the CME and only decelerated within the LASCO field of view.
Transition into the bow-shock regime most likely occurred at a
larger distance from the Sun.

\end{enumerate}

On the other hand, the analysis and considerations provide better
understanding what do microwaves show. The strong dependence of
the energy release on the magnetic field strength emphasizes
emissions from strong-field regions. They are additionally
emphasized by the strong dependence on the magnetic field of the
microwave emission. As section~\ref{s-flare} has demonstrated,
weaker microwave sources should not be neglected as they show
important parts of the flare configuration and its development.

This extreme event confirms the conclusions about the nature of
coronal shock waves drawn by \citet{Grechnev2011_I} from
observations of weaker events. The shock-wave nature of the
disturbances observed in this event is confirmed by the close
quantitative correspondence of their development to the expected
propagation of shock waves, whose excitation coincided in time and
space with strongest accelerations of eruptive flux ropes. Those
were near-surface ``EUV waves'', tracers of the type II bursts,
and the leading edge of the CME. These disturbances were
super-Alf{\'e}nic, at least, during some time after their
appearance, and only decelerated afterwards.

The scenario revealed for the event disagrees with the delayed
CME-driven bow-shock hypothesis: the shock developed much earlier
and could accelerate protons before the flare peak. The delayed
particle release time, which is sometimes inferred with respect to
the flare, can actually be due to the expansion of the CME
magnetic rope, where accelerated particles are trapped. They can
be released, when the rope reconnects with a streamer. Thus, the
late particle release time is not a consisting argument in favor
of exceptional shock-acceleration of solar energetic particles.

\bigskip

Acknowledgements. We bring our gratitude to the memory of Takeo
Kosugi, who devoted many years of his life to the progress of
Solar Physics doing his best for the solar microwave observations
in Nobeyama Radio Observatory and being one of major drivers for
the Yohkoh and Hinode solar missions.

We thank I.~Chertok, L.~Kashap\-ova, K.~Shibasaki, Y.~Hanaoka,
K.-L.~Klein, and Y.~Kubo for discussions and assistance. We are
indebted to the anonymous referee for useful remarks. We are
grateful to the Hinode team for all their efforts in the design,
development and operation of the mission. Hinode is an
international project supported by JAXA, NASA, PPARC and ESA. We
thank the colleagues from Nobeyama Solar Radio Observatory (NAOJ)
operating NoRH and NoRP, and the colleagues from the SSRT team
operating the Callisto spectrometer. We appreciate efforts of the
teams operating EIT, LASCO, and MDI on SOHO (ESA \& NASA);
GOES-12/SXI; HiRAS NICT, USAF RSTN, and Culgoora Solar
Observatory. We thank the team maintaining the SOHO LASCO CME
Catalog.

We acknowledge efforts of all researchers who contributed to study
of the 2006 December 13 event for valuable information even if
their conclusions have not been confirmed in our analysis. These
efforts certainly advance step by step better understanding of
solar phenomena, whose outcome can be significant for space
weather conditions.

This study was supported by the Russian Foundation of Basic
Research under grants 12-02-00037, 12-02-91161, 12-02-00173,
13-02-10009, 13-02-90472, 12-02-33110-mol-a-ved, and
12-02-31746-mol-a; the Program of basic research of the RAS
Presidium No.~22, and the Russian Ministry of Education and
Science under projects 8407 and 14.518.11.7047. N.M. was sponsored
by a Marie Curie International Research Staff Exchange Scheme
Fellowship within the 7th European Community Framework Programme.

\end{document}